\documentclass[a4paper]{article}

\usepackage[english]{babel}
\usepackage[utf8]{inputenc}
\usepackage{authblk} 
\usepackage[round]{natbib}
\setcitestyle{authoryear}

\usepackage{tabularx} 
\usepackage{booktabs} 
\usepackage{multirow}
\usepackage{amsmath}
\usepackage{amsthm} 
\theoremstyle{definition}
\newtheorem*{definition}{Definition}
\usepackage{graphicx}
\usepackage{amsfonts}
\usepackage{hyperref}
\usepackage{cleveref}
\usepackage[ruled,vlined,linesnumbered]{algorithm2e} 
\usepackage{subfig} 
\usepackage{tikz}
\usetikzlibrary{shapes,arrows}
\usetikzlibrary{backgrounds}
\usetikzlibrary{positioning}
\usepackage{listings}
\usepackage{color}
\usepackage[margin=1in]{geometry}

\definecolor{mygreen}{rgb}{0,0.6,0}
\definecolor{mygray}{rgb}{0.5,0.5,0.5}
\definecolor{mymauve}{rgb}{0.58,0,0.82}

\lstdefinelanguage{scala}{
  morekeywords={abstract,case,catch,class,def,%
    do,else,extends,false,final,finally,%
    for,if,implicit,import,match,mixin,%
    new,null,object,override,package,%
    private,protected,requires,return,sealed,%
    super,this,throw,trait,true,try,%
    type,val,var,while,with,yield},
  otherkeywords={=>,<-,<\%,<:,>:,\#,@},
  sensitive=true,
  morecomment=[l]{//},
  morecomment=[n]{/*}{*/},
  morestring=[b]",
  morestring=[b]',
  morestring=[b]"""
}

\newcommand\numberthis{\addtocounter{equation}{1}\tag{\theequation}}

\title{Functional probabilistic programming for scalable Bayesian modelling}

\author[1,2]{Jonathan Law}
\author[1,3]{Darren J. Wilkinson}
\affil[1]{School of Mathematics, Statistics \& Physics\\Newcastle University, U.K.}
\affil[2]{National Innovation Centre for Data}
\affil[3]{The Alan Turing Institute}

\begin{document}
\maketitle

\begin{abstract}
Bayesian inference involves the specification of a statistical model by a
statistician or practitioner, with careful thought about what each parameter
represents. This results in particularly interpretable models which can
be used to explain relationships present in the observed data. Bayesian models
are useful when an experiment has only a small number of observations and in applications where transparency of data driven decisions is important.
Traditionally, parameter inference in Bayesian statistics has involved constructing bespoke
MCMC (Markov chain Monte Carlo) schemes for each newly proposed statistical model. This results in
plausible models not being considered since efficient inference schemes are
challenging to develop or implement. Probabilistic programming aims to reduce
the barrier to performing Bayesian inference by developing a domain specific
language (DSL) for model specification which is decoupled from the parameter
inference algorithms. This paper introduces functional programming principles which can be used to develop an embedded probabilistic programming
language. Model inference can be carried out using any generic inference
algorithm. In this paper Hamiltonian Monte Carlo (HMC) is used, an efficient MCMC method requiring the gradient of the un-normalised
log-posterior, calculated using automatic differentiation.  The concepts
are illustrated using the Scala programming language.
\end{abstract}

A git repository containing runnable example code for the probabilistic models
and an illustrative implementation of forward and reverse mode automatic
differentiation is available at~\url{https://git.io/probprog}.

\section{Introduction}
\label{sec:intro}
Bayesian inference using probabilistic programming is becoming of interest to
large industry-backed artificial intelligence (AI) labs such as Google, Uber and
Stripe~\citep{tensorflowProbability, pyro2018, rainier2018}, despite deep learning continuing to solve practical problems with large amounts
of labelled data. Deep neural networks are typically used as a ``black
box'', and state of the art networks can involve many
parameters~\citep{huang2017densely}. Some dense convolutional neural networks
can have millions of parameters. The values of these parameters are impossible
to reason about independently and require a large amount of data to learn.

In a Bayesian model, the likelihood and prior distributions for each parameter
must is specified using the judgement of the statistician and/or subject matter experts. This results in an interpretable model which can be used to
explain relationships present in the data and perform accurate
predictions with appropriately quantified uncertainty. Bayesian methods are especially useful when an experiment has
only a small number of observations, for instance in an A/B test using a small
number of highly valuable customers or in a clinical trial. Bayesian methods
provide a natural encoding for uncertainty; if a small number of observations
are available to inform the likelihood then the posterior
distribution will reflect the prior. Specifying uncertainty honestly along with
predictions can allow practitioners to make informed decisions. Well specified
Bayesian models ensure industries such as insurance and finance, which price
products and shares based on mathematical and statistical models, can be
transparent in the decisions that they make.

Traditionally, parameter inference in Bayesian statistics has involved
constructing bespoke MCMC schemes for each newly proposed statistical model.
This results in new models being explored more slowly than practitioners would
like, sometimes leading to plausible models not being considered since
efficient inference schemes are challenging to develop or implement.
Probabilistic programming aims to reduce the barrier to performing Bayesian
inference by developing a domain specific language for model specification
which does not rely on a detailed understanding of the parameter inference
algorithms. This means that new models can be explored easily, without
significant programming effort.

There exist many domain specific languages (DSLs) for performing Bayesian inference, the most well known
within the statistical community are
BUGS~\citep{lunn2000winbugs} and JAGS~\citep{plummer2003jags} in which a simple
syntax can be used to specify complex hierarchical models. BUGS and JAGS both
use variants of Gibbs sampling to perform inference. Stan is another popular DSL which
uses similar syntax to BUGS and JAGS. Stan uses Hamiltonian Monte Carlo (HMC)
with automatic differentiation for efficient sampling~\citep{carpenter2016stan}
but is restricted to models with continuous parameters.
Stan also implements approximate variational inference schemes. 

Many existing frameworks for probabilistic programming, including BUGS, JAGS and
Stan use DSLs which are in turn compiled to lower-level languages.
For instance in a Stan script, arbitrary C++ can be written to define a model or
inference scheme, but in a C++ program the Stan DSL for defining a model cannot
be easily used. This is in contrast to an embedded DSL, where the DSL is written
directly in the target language and can be easily deployed inside a larger program.

TensorFlow is a program created at Google for deep learning and neural networks~\citep{tensorflow2015}.
Google has developed TensorFlow Probability (formally known as Edward), a
Python library for Bayesian inference utilising the TensorFlow architecture~\citep{tensorflowProbability}.
TensorFlow probability uses automatic differentiation and GPU acceleration from
TensorFlow for variational inference and HMC. Ride-sharing company Uber have developed a Python
library called Pyro~\citep{pyro2018} for probabilistic programming marrying deep
learning and Bayesian inference using PyTorch~\citep{paszke2017automatic} for
automatic differentiation and GPU acceleration. Pyro is implemented in Python, a
popular general purpose programming language. These probabilistic programming
languages can be easily deployed in existing Python code bases, however the
syntax for implementing a model is more verbose than that of BUGS, JAGS and Stan.
Hakaru~\citep{narayanan2016probabilistic} is an example of a probabilistic
programming language with both an embedded DSL written in Haskell and a more
accessible DSL written in the style of Python. This design choice was to
minimise the learning curve of those coming from languages similar to Python.

Within the theoretical computer science community, an effort has been made to
formalise the semantics of Bayesian inference in
category theory by defining a probability monad grounded in measure
theory~\citep{giry1982categorical,lawvere1962category}. A practical
implementation of Bayesian updating using the sampling based probability monad,
built upon the State monad with applications to robot
localisation and mapping is presented in~\cite{park2005probabilistic}.
A stochastic lambda calculus is introduced in~\cite{ramsey2002stochastic} along
with a review of the probability monad as presented in earlier literature, using
a Haskell-like notation. More recently, the use of the Giry monad as a foundation for probabilistic programming has been questioned. The issue is that the Giry monad is defined on the category of measurable spaces, but this category is not Cartesian closed. Since closure is desirable, alternative probability monads over Cartesian closed categories are being explored, such as quasi-Borel spaces~\citep{heunen2017convenient} and the Kantorivich monad~\citep{fritz2017probability}. The development of a composable probabilistic programming language using modern functional programming approaches is described in \cite{scibior2018functional}.

Rainier~\citep{rainier2018} is an embedded probabilistic programming language
written in Scala~\citep{odersky2004overview} developed by the online
payment processing company, Stripe. Rainier provides a compositional, 
monadic syntax for the specification of models. Each model has an associated compute
graph which can be used to calculate the un-normalised log-posterior and its
gradient using automatic differentiation. The log-posterior and its gradient is
then used for HMC~\citep{duane1987hybrid}. The monadic interface is familiar to
functional programmers and means that well researched and developed features of
existing functional programming languages can be used when developing probabilistic
programs. Crucially, models are are regular values within the host
programming language, which can be manipulated and composed using regular
language features. This is demonstrated in an example presented in this
paper, a random effects model (see section~\ref{sec:random-effects}). Each class
of the random effects model is modelled using an independent linear model, and the
linear model code from example~\ref{sec:lm-pp} is reused. This code reuse is
straightforward when using a well developed programming language such as Scala
and this is one of the main advantages of embedded probabilistic programming languages.

Scala is a programming language which runs on the Java Virtual Machine (JVM) and
provides support for functional
programming. Scala makes heavy use of function
composition, higher-order
functions and type classes to build large applications from simple functions.
These features allow programmers to build complex programs whilst maximising
code-reuse and minimising lines-of-code. This style of programming is a
departure from object-oriented or imperative programming and as such can have a steep learning curve for those coming from languages which emphasise those
approaches. This paper introduces the concepts required to understand monads in
functional programming and how they can be used to build an embedded DSL for
Bayesian statistical modelling.

Embedded DSLs are straightforward to write in functional languages using, for
example, free monads and generalised algebraic data
types~\citep{swierstra2008data}.  \cite{scibior2015practical} implements
an algebra in Haskell for operating on distributions using a Generalised
Algebraic Datatype and the Free Monad~\citep{scibior2015practical}. This allows the
implementation details of complex programs (in this case statistical inference
algorithms) to be decoupled from the model specification. Since the DSL is implemented in the same programming language, it
is straightforward to use within existing programs. The importance of building
DSLs for scientific computing is expanded upon in an article advocating the use
of Lisp in bioinformatics~\citep{khomtchouk2016strengths}. Lisps are a
collection of dialects of dynamically typed functional programming languages, which share some
similarities with functional programming in strongly typed functional languages such as Scala and Haskell.

\section{Category Theory}
\label{sec:category-theory}

This section introduces the aspects of category theory~\citep{awodey2010category,barr1990category}
required to build a probabilistic programming language. The concepts and their
uses are reified in the Scala language in section~\ref{sec:functional}. Category theory is concerned with describing abstract structures in a very
general way. For an introduction to
category theory aimed at programmers, see~\cite{milewski2018}.

\begin{definition}
  A \emph{category} $\mathcal{C}$ is a collection of objects and
  morphisms which go between them such that for the objects $X, Y, Z \in
  \mathcal{C}$ and morphisms $g: X \rightarrow Y$ and $h: Y \rightarrow Z$,

  \begin{enumerate}
  \item There must exist a morphism $f: X \rightarrow Z$ which is the composition of $g$ and
    $h$, $f = h \circ g$
  \item Each object in a category must have an identity morphism written as
    $\textrm{id}_X: X \rightarrow X$
  \item Composition of morphisms must be associative.
    For all $f: X \rightarrow Y$, $g: Y \rightarrow Z$ and $h: Z \rightarrow W$ then $h \circ (f \circ
    g) = (h \circ f) \circ g = h \circ f \circ g$
  \item For every morphism $g: X \rightarrow Y$ then $\textrm{id}_Y \circ g = g
    = g \circ \textrm{id}_X$.
  \end{enumerate}
\end{definition}

The collection of objects in a category $\mathcal{C}$ is often denoted $\textrm{obj}(\mathcal{C})$ and morphisms
$\textrm{hom}(\mathcal{C})$. The set of morphisms from $X$ to $Y$ is written as
$\textrm{hom}_{\mathcal{C}}(X, Y)$.

An example of a category is \textbf{Set}, the
category of sets: each set is an object and the morphisms are functions between
the sets, the identity morphism is the identity function. In functional
programming, the category most often under consideration is similar to the category of
sets: each object corresponds to a specific datatype and the morphisms represent functions between the types. 

\subsection{Functors}
\label{sec:functors-ct}

A functor is a structure preserving mapping from one category to
another.

\begin{definition}
  If $\mathcal{C}$ and $\mathcal{D}$ are categories, then a \emph{functor}
  $F:\mathcal{C} \rightarrow \mathcal{D}$ takes each object $X$ in
  $\mathcal{C}$ to an object $F(X)$ in the category $\mathcal{D}$ and each
  morphism, $f:X \rightarrow Y$ in $\mathcal{C}$ to a morphism in $\mathcal{D}$, 
  $F(f): F(X) \rightarrow F(Y)$ such that

  \begin{enumerate}
  \item $F(\textrm{id}_X) = \textrm{id}_{F(X)}$ for each $X$ in $\mathcal{C}$
  \item $F(g \circ f) = F(g) \circ F(f)$ for all morphisms, $f$ and $g$ in $\mathcal{C}$.
  \end{enumerate}
\end{definition}
Since functional programming is mainly concerned with the category \textbf{Set},
functors are typically endofunctors, $F: \textbf{Set} \rightarrow
\textbf{Set}$. An endofunctor is a functor from a category to itself, $F: \mathcal{C} \rightarrow \mathcal{C}$. Note that every category $\mathcal{C}$ has an associated identity endofunctor, $\textrm{Id}_\mathcal{C}: \mathcal{C}\rightarrow\mathcal{C}$ mapping all objects and morphisms to themselves.

\subsection{Natural Transformation}
\label{sec:nt}

A functor is a morphism between two categories (which itself has morphisms
between its objects). A \emph{natural transformation} is a morphism between
functors which preserves the structure of the categories.

\begin{definition}
  If $F: \mathcal{C}
  \rightarrow \mathcal{D}$ and $G: \mathcal{C}
  \rightarrow \mathcal{D}$ are both functors between the categories $\mathcal{C}$
  and $\mathcal{D}$ then a \emph{natural transformation} $\alpha: F \Rightarrow G$
  is a family of morphisms such that
  \begin{enumerate}
  \item  For all $X \in \mathcal{C}$ then $\alpha_X: F(X) \rightarrow G(X)$ is a morphism in
    $\mathcal{D}$ called the component of $\alpha$ at $X$
  \item For each morphism $f \in \mathcal{C}$, $f: X \rightarrow Y$ then
    $\alpha_Y \circ F(f) = G(f) \circ \alpha_X$.
  \end{enumerate}
\end{definition}

\begin{figure}
  \centering
  \includegraphics[width=0.4\textwidth]{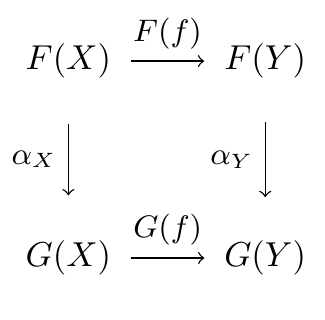}
  \caption{Commutative diagram of the natural transformation}
  \label{fig:natural-trans}
\end{figure}

Figure~\ref{fig:natural-trans} shows a commutative diagram of the second natural
transformation law. Following the arrows of the commutative diagram for any path from any
given start to end point must be equivalent.

\subsection{Monads}
\label{sec:monads-ct}

\begin{figure}
  \centering
  \subfloat[]{
    \includegraphics[width=0.4\textwidth]{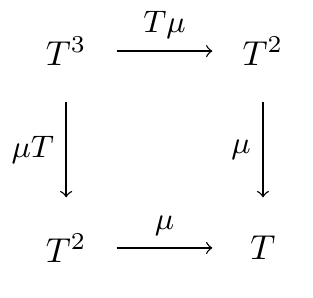}}\qquad
  \subfloat[]{\includegraphics[width=0.4\textwidth]{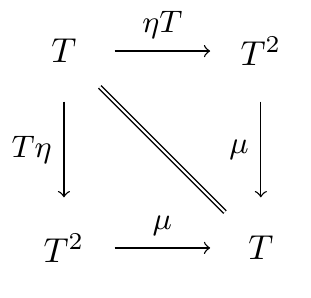}}
  \caption{(a) Commutative diagram expressing the first condition of the
    monad laws (b) Commutative diagram expressing the second condition of
    the monad laws}
  \label{fig:monad-laws}
\end{figure}

\begin{definition}
  A monad on the category $\mathcal{C}$ is an endofunctor, $T: \mathcal{C}
  \rightarrow \mathcal{C}$ with two natural transformations, $\eta:
  \textrm{Id}_{c} \rightarrow T$ called \textbf{unit} and $\mu: T \circ T
  \rightarrow T$ called \textbf{multiplication} (or \textbf{counit}) such that 

  \begin{enumerate}
  \item $\mu \circ T \mu = \mu \circ \mu T$ 
  \item $\mu \circ T\eta = \mu \circ \eta T = \textrm{Id}_T$
  \end{enumerate}
\end{definition}
$T\mu: T(T \circ T) \rightarrow T$ is a natural transformation defined by taking
the natural transformation $\mu$ and the endofunctor, $T: \mathcal{C} \rightarrow
\mathcal{C}$ then $(T\mu)_C = T\mu_C$. The natural transformation $\mu T$ is a
similar transformation with $T$ composed from the right, $\mu T: (T \circ T) T
\rightarrow T$. Figure~\ref{fig:monad-laws} shows commutative diagrams of the monad laws.

\section{Principles of Functional Programming}
\label{sec:functional}

In functional programming, a program is comprised of many small functions each
of which are easy to test and reason about in isolation. These small functions
are composed together to form a larger program, which in practice is more
difficult to reason about and test. The programmer gains confidence in the
correctness of the program by ensuring each individual function in the
composition is correct. This section provides an introduction to the functional
programming principles required to build a probabilistic programming language
guided by category
theory, with examples in Scala. For a more comprehensive introduction to
functional programming in Scala, see~\cite{chiusano2014functional}. 

\subsection{Referential Transparency and Pure Functions}
\label{sec:refer-transp}

\emph{Referential transparency} is a fundamental principle of functional
programming. A function is referentially transparent if it produces the same
output value for the same function parameters upon repeated evaluation, and
produces no side effects. This
enables the programmer to directly replace any referentially transparent
function with the value of that function without changing the output of the
program. This is called reasoning by substitution. A function which is
referentially transparent is often called deterministic. 

A \emph{pure function} is a function which is referentially transparent,
inculpable and total. A function is total if it returns a value (not an
exception) for every possible input. Pure functions are more straightforward to
test, reason about and reuse by composing with other pure functions. A function is called \emph{inculpable} if it is
free from side effects. 

Functions which have \emph{side effects} are not referentially transparent. An
example of a side effect is printing to the screen, reading from a file or
database, accepting input from the user or mutating a non-local variable. Some or
all of these actions are needed for a program to be useful. Functional
programming makes side effects explicit and good functional design ensures that
side effects are only performed when needed, typically at the edges of a program.

An example of a common side effect in statistical programming is generating a
random number. In Scala (which is not a \emph{pure} functional language) a random double between 0 and 1 can be generated using
\lstinline{scala.util.Random.nextDouble()}. When calling this function it is
impossible to know which random number will appear next without knowing the
current state of the random number generator. Additionally, when this function
is called multiple times in a program it will return different values. This is
due to the fact that the function has the side effect of mutating the global
random number state. The state
monad can be used to implement a purely functional random number generator and this
is expanded upon in section~\ref{sec:functional-prng}.

\subsection{Static Types}
\label{sec:static-types}

Functions can accept a range of parameters and each of these parameters has a
type. For instance the function $f(x) = x + 1$ is a mathematical function which
adds one to its parameter, $x$. In this case, the parameter type is implicitly a
numeric type as the method $+$ is associated with numbers. The return type of
the function $f$ is also a numeric type. Types can be either inferred (as in the
function $f$) or they can be made explicit. The example Scala code below shows a the function $f$ written with an
explicit type declaration of \lstinline{Int => Int}. 

\begin{lstlisting}
val f: Int => Int = x => x + 1
\end{lstlisting}
If this function was applied to a \lstinline{String} data-type, the program
would not compile. In a language with dynamic types, the program would run and
possibly try to convert the \lstinline{String} into an \lstinline{Int} this can
lead to runtime errors which are typically difficult to debug as they can
potentially go unnoticed.

Static types are not exclusive to functional programming, nor are they required.
However, programs using static types are typically safer and easier to reason about; Scala
and Haskell are examples of functional programming languages with static types. However, both Scala and Haskell have \emph{type inference}, which means that explicit type annotations are not required in situations where the compiler can unambiguously resolve all types; this provides most of the advantages of static typing without the verbosity associated with statically typed languages lacking type inference.

\subsection{Immutable Data}
\label{sec:immutable-data}

\emph{Immutable data} is an important aspect of functional programming. An immutable value or data structure cannot be modified in place. This is a strict requirement which ultimately makes code easier to reason about. However this means
that imperative programming constructs, such as for-loops and while-loops can
not be used since, for instance, the index of the loop can not be advanced (by
mutating the variable representing the index) as would be required in a for-loop.

\subsection{Higher Kinded Types}
\label{sec:kinds}

Collection types such as a singly linked lists, represented by \lstinline{List},
are sometimes referred to as a type constructors (or data constructors) since
they are used to construct concrete types such as \lstinline{List[Int]} or \lstinline{List[Double]}. \lstinline{List} is
similar to other collection types such as \lstinline{Vector} or
\lstinline{Option}. Vectors and lists are collections of zero or more
elements. Whereas an \lstinline{Option} is a container for turning a partial function into a total
function by explicitly representing failure cases using \lstinline{None}.
\lstinline{Option} can be thought of as a collection which can be either empty
(\lstinline{None}) or contain exactly one element.

The implementations of \lstinline{List} and \lstinline{Vector} are
different and both have distinct performance characteristics. A singly linked
list provides $O(1)$ access to the head of the list but $O(n)$ access time to a
specific element. Whereas \lstinline{Vector} in Scala is implemented as a
trie~\citep{Fredkin1960} with a branching factor of 32. This means that random
access to elements in a \lstinline{Vector} is faster than in a \lstinline{List}.
Consequently, concrete collections should be appropriately chosen for a given purpose.

These collections (and other type constructors) share common abstractions
which can be used to unify them and then write code which is polymorphic not
just in the type, but in the type constructor. This allows for greater code
re-use. Higher kinded types provide powerful abstractions leading to cleaner,
more elegant code.
Very few mainstream programming languages include the necessary language support for using higher kinded types to safely abstract over generic type constructors, with Scala and Haskell arguably being the best known examples of languages which do.

\subsection{Higher Order Functions}
\label{sec:hof}

In programming languages which have functions as first class citizens, functions
can be passed as an argument to other functions. Functions which accept other
functions as arguments or return functions are called \emph{higher order
  functions}. 

A common operation performed on collections using a for-loop is to apply a
function, \lstinline{f}, to each element in the collection. The \lstinline{map}
function can be used to apply a function to a value in a context, and a context,
as in this case, can be a collection. The signature of the \lstinline{map} function is 

\begin{lstlisting}
def map[F[_], A, B](fa: F[A])(f: A => B): F[B]
\end{lstlisting}
\lstinline{map} is a higher order function which accepts a function,
\lstinline{f: A => B} and applies it to each element of \lstinline{F}. The term
\lstinline{F} is used to stand for any construct which can be mapped over using
a function, this construct is known as a Functor (see
section~\ref{sec:functors}). The relationship with the category theoretic
functor is made clearer if the \lstinline{map} is restated such that it
transforms the function \lstinline{f: A => B} into \lstinline{g: F[A]  => F[B]}.

Another common operation for collections is a reduction, and this can be achieved by
applying a binary function to each pair of elements of the collection in order from
either the left or right. A higher order function which reduces values of the
list from the left starting with an initial seed value is \lstinline{foldLeft},
the signature of \lstinline{foldLeft} is

\begin{lstlisting}
def foldLeft[F[_], A, B](fa: F[A])(b: B)(f: (B, A) => B): B
\end{lstlisting}
\lstinline{map} and \lstinline{foldLeft} are implemented using recursion, but in
order to avoid stack overflows they are implemented using tail-calls which are
straightforward for the compiler to optimise. In contrast to imperative
for-loops these higher order functions require no auxiliary variables such as
counters. They are typically shorter and express the intent of the programmer more
clearly.

\subsection{Functors}
\label{sec:functors}

A functor represents a type constructor which can be
\lstinline{map}ped over. A functor can be used to lift a function between two
types into a new function between two types in a context, where the context
refers to the type constructor. The signature of the \lstinline{map} function for each collection \lstinline{List}, \lstinline{Option} or \lstinline{Vector} can be written as follows:

\begin{lstlisting}
def map[A, B](fa: List[A])(f: A => B): List[B]
def map[A, B](fa: Option[A])(f: A => B): Option[B]
def map[A, B](fa: Vector[A])(f: A => B): Vector[B]
\end{lstlisting}
The function \lstinline{f: A => B} is lifted into the context of \lstinline{List}, \lstinline{Option} and \lstinline{Vector} respectively. The higher order function \lstinline{map} allows \lstinline{f} to operate on an element in a context. The \lstinline{map} functions defined above are identical apart from the type constructor, the type constructor can thus be abstracted over into the functor type class:

\begin{lstlisting}
trait Functor[F[_]] {
  def map[A, B](fa: F[A])(f: A => B): F[B]
}
\end{lstlisting}
The type constructor, denoted as \lstinline{F[_]}, has been abstracted over
using a \emph{higher kinded type}, here a functor. Now functions can be defined with
a functor context bound on the type constructor:

\begin{lstlisting}
def addOneGen[F[_]: Functor, A: Numeric](fa: F[A]): F[A] = {
  map(fa)(a => a + 1)
}
\end{lstlisting}
This function can be applied to any data constructor which has a functor
instance and which contains Numeric values. In order
to use the function, \lstinline{addOneGen} with a concrete collection type, a
functor instance must be defined by defining the \lstinline{map} function. The
Scala library Cats\footnote{https://github.com/typelevel/cats} provides type class instances for many of the built in Scala types constructors. 

\subsection{Applicatives} 
\label{sec:app}

The data constructors considered so far are also applicatives~\citep{mcbride2008applicative}. An applicative
is a special functor which has a \lstinline{pure} method which lifts a value into the context of the
Applicative and a \lstinline{ap} which can be used to apply a function to a
value in the Applicative context (the \lstinline{F[_]})

\begin{lstlisting}
trait Applicative[F[_]] extends Functor[F] {
  def pure[A](a: A): F[A]
  def ap[A, B](fa: F[A])(f: F[A => B]): F[B]
}
\end{lstlisting}
To define \lstinline{pure}, use the type constructors for the collections:

\begin{lstlisting}
def pure[A](x: A): List[A] = x :: Nil
def pure[A](x: A): Option[A] = Some(x)
def pure[A](x: A): Vector[A] = Vector(x)
\end{lstlisting}
Where \lstinline{::} is used as an infix operator which can be used to construct
a singly-linked list by prepending a single element on the left to a list on the
right. In Cartesian closed categories (CCCs), like Set, \lstinline{ap} can be defined in
terms of \lstinline{zip} and \lstinline{map}. In category theory an applicative defined in terms of \lstinline{ap} is a \emph{lax closed functor} and an applicative
defined in terms of \lstinline{zip} is a \emph{lax monoidal functor}. In CCCs, these are equivalent. As such, the
applicative typeclass can be defined in terms of the functions \lstinline{zip},
\lstinline{map} and \lstinline{pure}. The type signature of \lstinline{zip} is
given by

\begin{lstlisting}
def zip[A, B](fa: F[A], fb: F[B]): F[(A, B)]
\end{lstlisting}
Then \lstinline{ap} can be derived

\begin{lstlisting}
def ap[A, B](fa: F[A])(f: F[A => B]): F[B] = 
  map(zip(f, fa)){ case (fi, a) => fi(a) }
\end{lstlisting}
The definition of \lstinline{zip} for the \lstinline{List} collection can be defined
recursively as follows, 

\begin{lstlisting}
def zip[A, B](fa: List[A], fb: List[B]): List[(A, B)] = (fa, fb) match {
  case (Nil, lb) => Nil
  case (la, Nil) => Nil
  case (a :: ta, b :: tb) => (a, b) :: zip(ta, tb)
}
\end{lstlisting}
If the lists are different lengths, \lstinline{zip} will return a list of tuples
the same length as the shortest list. 

\subsection{Monads}
\label{sec:monad}

The collections considered so far are actually monads. A monad represents a value in
a context which can be manipulated in a consistent way~\citep{wadler1995monads}.
Monads are integral to safe purely functional programming and are used to
encapsulate unsafe program behaviour such as IO (input output), async programs,
random number generators and partial functions. Suitable monads used in Scala for these unsafe
behaviours include the \lstinline{IO} monad (implemented in Cats
Effect\footnote{https://typelevel.org/cats-effect/}), \lstinline{Future}, \lstinline{Rand} and \lstinline{Option} respectively. These monads have unsafe
methods which can be used to access the contents of the monad and hence perform
the side effect. The unsafe access method is only performed at the end of the
program when the side effect is desired, until then combinations of
\lstinline{map} and \lstinline{flatMap} are used to manipulate the values inside
of the monadic context. 

To define a monad in Scala, simply define the two natural transformations
required. The natural transformation corresponding to $\eta: \textrm{Id}_C
\rightarrow T$ is \lstinline{pure}, inherited from the \lstinline{Applicative}
type class. The second natural transformation, $\mu: T \circ T \rightarrow T$, is
referred to as \lstinline{join}

\begin{lstlisting}
trait Monad[F[_]] extends Applicative[F[_]] {
  def join[A](fa: F[F[A]]): F[A]
}
\end{lstlisting}
Typically, in functional programming monads
are defined in terms \lstinline{flatMap}

\begin{lstlisting}
def flatMap(fa: F[A])(f: A => F[B]) =
  join(map(fa)(f))
\end{lstlisting}
which is a map followed by a \lstinline{join} or flatten. The monad can
equivalently be defined in terms of \lstinline{flatMap} and \lstinline{pure}

\begin{lstlisting}
trait Monad[F[_]] extends Functor[F[_]] {
  def flatMap[A, B](fa: F[A])(f: A => F[B]): F[B]
  def pure[A](x: A): F[A]
}
\end{lstlisting}
Note that \lstinline{map} can be written in terms of \lstinline{flatMap} and
\lstinline{pure} and \lstinline{ap} can be written in terms of
\lstinline{flatMap} and \lstinline{map}. This means when a monad instance is
defined for a given type, only \lstinline{flatMap} and \lstinline{pure} must be
defined and the methods from functor and applicative are available. Hence all
monads in CCCs are applicative functors, though the converse is not true. The ubiquity of monads in functional programming derives from the power of \lstinline{flatMap}, which enables composition of functions returning values in a monadic context. Explicitly, functions \lstinline{f: X => F[Y]} and \lstinline{g: Y => F[Z]} don't compose directly, since the types don't align, but for a monad \lstinline{F[_]}, they can be composed using \lstinline{flatMap} to give a function \lstinline{X => F[Z]}. The ability to chain together such monadic functions turns out to be exceptionally useful.

\lstinline{Option} is a monad which can be used to turn a partial function into
a total-function by capturing error states in the type information. Consider several functions which could fail:
\begin{lstlisting}
def sqrt(x: Double): Option[Double] =
  if (x < 0) None else Some(math.sqrt(x))
def log(x: Double): Option[Double] =
  if (x <= 0) None else Some(math.log(x))
\end{lstlisting}
In order to compose these two functions, a \lstinline{flatMap} can be used:
\begin{lstlisting}
def composed(x: Double): Option[Double] =
  sqrt(x).flatMap(y => log(y))
\end{lstlisting}
The same principal can be extended to more than two applications:
\begin{lstlisting}
def composeMore(x: Double): Option[Double] =
  sqrt(x).flatMap(y => log(y)).flatMap(z => log(z))
\end{lstlisting}
This can be visually improved using a \lstinline{for}-comprehension which is 
syntactic-sugar for applications of \lstinline{flatMap} and \lstinline{map}:
\begin{lstlisting}
def composeFor(x: Double): Option[Double] = for {
  y <- sqrt(x)
  z <- log(y)
  res <- log(z) 
} yield res
\end{lstlisting}
The functions returning an \lstinline{Option} (or any monad) are placed on the right side of
the arrow \lstinline{<-} extracting the value from the monadic context. The
values on the left represent the value inside of the monadic context and can be
referenced in the same for-comprehension. The function \lstinline{composeFor} is
equivalent to \lstinline{composeMore}. The for-comprehension is ``desugared'' by the compiler
into a chain of \lstinline{flatMap}s and \lstinline{map}s in a very early compiler pass. The equivalent to a \lstinline{for}-comprehension in
Haskell is known as \verb$do$-notation.

In general, \emph{different} monads do not compose~\citep{jones1993composing}, hence
functions with different monadic contexts can not be mixed inside of a
for-comprehension. However, different contexts are often required, and for certain
combinations, monad transformers can be used which combine monadic
contexts~\citep{liang1995monad}. Alternatively, extensible effects and the Freer monad can be used
for combining effects~\citep{kiselyov2013extensible}.

\subsection{Monadic Collections}
\label{sec:monad-collections}

Modern hardware is parallel, with most laptops containing CPUs with multiple
cores. The sequential collection types considered so far, \lstinline{List} and
\lstinline{Vector} do not take advantage of multithreading. In Scala there are
parallel collections which share the same monadic interface, and this enables
functions to be written in a generic way and easily parallelised to take
advantage of environments with multiple CPU cores.

Any collection can be transformed into a parallel collection, by calling
\lstinline{.par}. Higher order functions defined on the parallel
collections will run in parallel. \lstinline{map} and \lstinline{filter} are
naturally parallel and are evaluated by splitting the collection into roughly
equal sized chunks
and evaluating a sequential \lstinline{map} or
\lstinline{filter} on each partition using different threads. Reduction
operations need more careful consideration. \lstinline{foldLeft} and
\lstinline{foldRight} are inherently sequential, however \lstinline{reduce} can
be implemented using an associative binary reduction function and then the
reduction can be performed in parallel using a parallel tree reduction.

Consider evaluating the likelihood of a linear model, the specification of the
model is presented in example~\eqref{sec:lm-pp}. The likelihood can be written as:

\begin{equation*}
  p(y|\psi) = (2\pi \sigma^2)^{-\frac{N}{2}}\exp\left\{-\frac{1}{2\sigma^2}\sum_{i=1}^N(y_i - (\beta_0 + \beta_1 x_i))^2\right\}
\end{equation*}
where $\psi = \{\beta_0, \beta_1,\sigma\}$. The log-likelihood can then be written as:

\begin{equation*}
  \log p(y|\psi) = -\frac{N}{2}\log(2\pi\sigma^2)-\frac{1}{2\sigma^2}\sum_{i=1}^N(y_i - (\beta_0 + \beta_1 x_i))^2
\end{equation*}
This log-likelihood can be evaluated in Scala using a \lstinline{map} followed
by a \lstinline{reduce}

\begin{lstlisting}
ys.
  map { case (x, y) => Gaussian(beta0 + beta1 * x, sigma).logPdf(y) }.
  reduce(_ + _)
\end{lstlisting}
Where \lstinline{ys: F[(Double, Double)]} is a collection of tuples containing the dependent and independent variables. This code is completely agnostic to whether the collection being operated on is serial or parallel. If \lstinline{ys} is a serial collection, then \lstinline{map} and \lstinline{reduce} will execute serially on a single core. However, if \lstinline{ys} is a parallel collection, then \lstinline{map} and \lstinline{reduce} will execute on the collection in parallel on all available cores.
In practice, the parallel reduction will only speed up processing significantly if the
collection of data is large, or the individual likelihood functions are expensive
to evaluate.

If the data contained in the collection is too large to fit into the memory of a
single computer, then higher order functions such as map and reduce can be used for
distributed computing~\citep{dean2008mapreduce}. This pattern was named
MapReduce after the familiar higher order functions. MapReduce is commonly used
when a large amount of input data which
does not fit in the memory of a single node requires processing. The large dataset is partitioned onto several worker nodes, orchestrated by a single
master node. During the map stage, each worker node then performs a
function independently on each element of the data. In the reduce stage a binary
operator is used to combine the elements. The map and reduce steps can be performed by multiple nodes in parallel, meaning computationally intensive tasks
can be sped up by increasing the number of worker nodes. In practice the speed
of computation does not increase linearly, because of the additional cost of messaging
between nodes. 

MapReduce is a general pattern which can be used to implement many algorithms,
including the page rank algorithm used by Google to rank search results~\citep{brin1998anatomy}.
MapReduce aims to make distributed programming easier by removing the
possibility of race-conditions. Listing~\ref{lst:map-reduce} shows a word count
implementation written in the
MapReduce style. The \lstinline{map} stage assigns the number one to each word
in the corpus of documents. \lstinline{groupBy} is used to group the words into
key-value pairs where the key
is a single word and the value is a collection of tuples containing the
same word and the number one. The reduce stage is implemented using \lstinline{map} and \lstinline{reduce}, the
\lstinline{map} takes each tuple containing each word and collection and
performs the binary reduction function \lstinline{+} to add the numbers
associated with each word. The \lstinline{reduce} function is a
\lstinline{fold} without an initial value, as such it can only be performed on a
non-empty collection.

\begin{lstlisting}[caption={Illustrative MapReduce program written using Scala},label={lst:map-reduce}]
val wordCount: Map[String, Int] = words.
  map(w => (w -> 1)).
  toList.
  groupBy(_._1).
  map { case (w, ws) => (w, ws.map(_._2).reduce(_ + _)) }
\end{lstlisting}
Apache Spark~\citep{spark2018} is a sophisticated platform for distributed
computing for big data. Spark utilises a high level functional interface
written in Scala allowing parallel processing for large datasets and streaming
data. Spark features APIs (application programming interfaces) for Scala, Python
and Java, and each API utilises higher order functions which can be used to
manipulate data across a cluster of compute nodes. Spark primarily stores data
in main memory of compute nodes, whereas Hadoop~\citep{hadoop2018} (a map reduce implementation)
stores the data on the hard disk drive (HDD) or solid-state drive (SSD) of the
worker nodes. This means that Spark is typically orders of magnitude faster for
the same tasks. Apache Spark has additional capabilities, including batch
processing of streaming data and a library for machine learning called MLlib. The main collection type in Apache spark is the resilient distributed dataset
(RDD) which is an immutable distributed monadic collection partitioned onto different
nodes in the Spark cluster. The RDD can be operated on using the higher order
functions familiar to users of the built in Scala collections. 

\subsection{Purely Functional Pseudo Random Number Generator}
\label{sec:functional-prng}

The \lstinline{State} monad can be used to keep track of internal state which is
cumbersome to pass around as an argument. The state monad is a constructor with a function from the
current state to a tuple containing a value and an updated state: 

\begin{lstlisting}
case class State[S, A](run: S => (S, A))
\end{lstlisting}
Since it is a monad, there are implementations of \lstinline{flatMap} and \lstinline{pure}:

\begin{lstlisting}
def flatMap[B](f: A => State[S, B]): State[S, B] = State { x =>
  val (st, a) = run(x)
  f(a).run(st)
}
def pure[B](a: B) = State(st => (st, a))
\end{lstlisting}
The \lstinline{State} monad can be used to implement a purely-functional
pseudo-random number generator (PRNG) by threading an PRNG state through
multiple function calls.

Sequences of apparently random numbers are generated by mutating an internal PRNG
state each time a function returning a random number is
evaluated. Each time a pseudo-random number is asked for a different value is
returned and the internal PRNG state is mutated

\begin{lstlisting}
def randNaive = scala.util.Random.nextDouble()
randNaive
// 0.26167644649515154
randNaive
// 0.4732297985296451
\end{lstlisting}
This function violates referential transparency (note that a \lstinline{def} is needed here rather than a \lstinline{val} since \lstinline{val} is eagerly evaluated) since each time \lstinline{randNaive} returns different values at each invocation. In order to develop a purely functional random number generator the internal state can be passed around explicitly. One of the simplest implementations of a pseudo random number generator is a linear congruential generator. This generator is defined by a recursive relation

\begin{equation}
  \label{eq:18}
  X_{n+1} = (aX_n + c) \operatorname{mod} m
\end{equation}
$m > 0$ is the modulus, $0 \leq c < m$ is the increment and $0 < a < m$ is the
multiplier and the initial seed is $0 \leq X_0 < m$. It is challenging to choose
the parameters $a, c$ and $m$ to produce a long apparently random chain, a
problem which is not considered here. The generator can be implemented in Scala as follows:

\begin{lstlisting}
def linearGenerator(a: Long, c: Long, state: Long) =
  a * state + c
def toDouble(x: Long) =
  (x >>> 11) * math.pow(2, -53)
def myGenerator(state: Long): (Long, Double) = {
  val newState = linearGenerator(6364136223846793005L,
 1442695040888963407L, state)
  (newState, toDouble(newState))
}
\end{lstlisting}
The linear generator is initialised without a modulus, which is equivalent to
the modulus being $2^{64}$ (since a \lstinline{Long} is 64 bits in Scala). The function \lstinline{toDouble}
performs a logical right bit-shift by 11 places, then divides by $2^{53}$ (since $53 =
64 - 11$).
The right shift of the \lstinline{Long} value by 11 bits and the subsequent
division ensures there is no numeric overflow, and avoids sign issues that would arise if the first bit was set. This has the effect of
transforming the seed to be a \lstinline{Double} between zero and one. In order
to generate a new number \lstinline{myGenerator} must be given an initial value,
which is the initial state of the of the random number generator: 

\begin{lstlisting}
myGenerator(13515670)
// (3730281199248434349,0.2022189490103492)
\end{lstlisting}
This function returns the next state on the left of the tuple and a ``random''
number between zero and one. The function can be initialised from a ``random''
or chosen seed. In practice, pseudo-random number generators are often more complex
than this; a well used PRNG with a long non-repeating chain is the Mersenne
Twister algorithm~\citep{matsumoto1998mersenne}.  

A function to draw a pseudo-random number can be now be written as:
\begin{lstlisting}
type Rand[A] = State[Long, A]
def randDouble: Rand[Double] = 
  State(myGenerator)
\end{lstlisting}
Then in order to sample several numbers, \lstinline{flatMap} can be used (hidden in a \lstinline{for}-comprehension):

\begin{lstlisting}
val res = for {
  firstNumber <- randDouble
  nextNumber <- randDouble
} yield (firstNumber, nextNumber)
println(res.run(13515670))
// (2060103227662993080,(0.2022189490103492,0.11167841974883075))
\end{lstlisting}
The State monad passes along the state to the next invocation of
\lstinline{randDouble}. In order to extract the pseudo-random numbers from
the state monad, the function \lstinline{run} is called with a starting seed.
The result is on the right of the tuple with the next state on the left. This
construct forms a referentially transparent PRNG. Note also, that the same
initial seed is used as in the previous example and hence the same pseudo-random
number is produced. The next number however is different, showing that the PRNG
state is passed along by the \lstinline{State} monad when \lstinline{randDouble}
is called for the second time.

Since \lstinline{State} is a
functor, \lstinline{map} can be used to transform the output of the function
\lstinline{randDouble} in order to build other
random generators:

\begin{lstlisting}
def randInt(from: Int, to: Int): Rand[Int] =
  randDouble.map(x => (x * (to - from) + from).toInt)
def randBool: Rand[Boolean] =
  randDouble.map(_ > 0.5)
\end{lstlisting}
The State monad can be used to
define a \lstinline{Rand} monad using other algorithms which
generate uniformly distributed random variables, and these can be transformed
appropriately to generate values distributed according to other useful
distributions. For example the Box-Muller transform is an efficient method of
generating standard Normal random numbers~\citep{box1958}.

\section{Probability Monads}
\label{sec:prob-monad}

In order to define a probabilistic program using the abstractions introduced so
far, it remains to show that probability measures form a monad. Monads are well
understood and ubiquitous in functional programming and programming language
theory, hence have well supported language features such as the
\verb$for$-comprehension in Scala and \verb$do$-notation in Haskell. Their uniformity enables
programmers to manipulate monadic values in a consistent way using
\lstinline{map} and \lstinline{flatMap} independent of the
monadic context. This makes defining probabilistic programs straightforward
using powerful, well-developed functional programming languages with support for
monads.

The foundations of the probability monad were first developed
by~\cite{lawvere1962category}, later extended in~\cite{giry1982categorical}. A
review of the Giry monad is presented in~\cite{ramsey2002stochastic}.
The Giry monad is defined on the category of measurable spaces, \textbf{Meas}. The Giry monad maps each measurable space $X$ to the space of of all probability measures on $X$, which we will denote $P(X)$. The monad is characterised by the two natural transformations $\eta$ and $\mu$. The unit $\eta_X: X \rightarrow P(X)$ must map each $x\in X$ to a probability measure on $X$, and does so in the obvious way by mapping to the Dirac measure, $\eta_X(x) = \delta_x$. The multiplication $\mu_X: P(P(X)) \rightarrow P(X)$ must flatten a probability measure over probability measures on $X$, down to a probability measure on $X$, and does so in the obvious way, by marginalisation (via Lebesgue integration). Then the monadic bind operation (\verb$flatMap$ in Scala) can be interpreted as the law of total probability. Explicitly, for probability kernels (conditional distributions) $f: X\rightarrow P(Y)$ and $g: Y\rightarrow P(Z)$ we bind them as $g \bullet f: X\rightarrow P(Z)$ via
\[
g \bullet f = \int_Y g(y)\, \textrm{d}f(x).
\]

One issue with the Giry monad is that it is defined on
\textbf{Meas}, which is not Cartesian-closed. This is not ideal for
any kind of higher-order probabilistic programming language, since it
prevents convenient use of probability distributions over
functions. It is also awkward for formalising the semantics of monadic
probabilistic programming languages developed as embedded DSLs in
functional programming languages with support for monads,
since there will typically be an implicit assumption that all monads
are defined over CCCs.

Measurable spaces are very rich, and
include many pathological cases which are not of significant practical
interest, yet cause substantial technical difficulties. So one
approach to ``fixing'' the problem is to restrict attention to spaces
of more direct relevance, in a principled way. This has led
to the development of a probability monad based on Quasi-Borel spaces,
which form a CCC, as explored in~\cite{heunen2017convenient}. Other
approaches make a more radical departure from conventional
measure-theoretic probability, such as the Kantorovich monad, defined
on metric spaces~\citep{fritz2017probability,jacobs2018monads}. The use of probability monads as a foundation for Bayesian inference is explored in~\cite{culbertson2014categorical}.

This theoretical underpinning paves the way for developing monadic
probabilistic programming DSLs in functional programming languages
with minimal effort beyond specifying the two natural transformations
required for the distribution to form a monad (see
section~\ref{sec:monad}). Although the formal connection with the
theory of probability monads is important for establishing rigorous
semantics for a given probabilistic programming language~\citep{staton2017commutative,scibior2018denotational}, the practical
implications associated with picking a specific probability monad
formalism for the development of actual implementations are limited~\citep{scibior2015practical,scibior2018functional}.

\section{Hamiltonian Monte Carlo}
\label{sec:hmc}

One of the aims of probabilistic programming is to separate the details of
inference algorithms from the specification of statistical models. This allows
for rapid exploration of different models without having to devise new
inference algorithms. Metropolis-Hastings (MH) is an MCMC algorithm with convergence
guarantees and if run for long enough will give samples from the stationary
distribution corresponding to the posterior
distribution of interest in a Bayesian statistical model~\citep{metropolis1953,
  hastings1970}. The MH algorithm uses a parameter proposal
distribution which requires tuning in order to achieve the optimal acceptance
rate of around 0.234~\citep{roberts1997weak}. Adaptive methods can be used to tune the proposal
distribution automatically~\citep{atchade2005adaptive}. Even optimal tuning of
the proposal distribution does not take into account the geometry of the
posterior distribution, which is contained in the gradient of the posterior. A
more efficient inference algorithm is desirable for general probabilistic programming and Bayesian inference.

Hamiltonian Monte Carlo utilises the gradient of the un-normalised log-posterior
to more efficiently explore the posterior distribution. The intuition for HMC is developed from Hamiltonian dynamics. Hamilton's equations are written as:

\begin{align*}
  \frac{\mathrm{d}\boldsymbol{p}}{\mathrm{d}t} &= -\frac{\partial \mathcal{H}}{\partial \boldsymbol{q}}, \\
\frac{\mathrm{d}\boldsymbol{q}}{\mathrm{d}t} &= +\frac{\partial \mathcal{H}}{\partial \boldsymbol{p}},
\end{align*}
where $\boldsymbol{q}$ represents a particle position and $\boldsymbol{p}$ is the momentum of the particle. The Hamiltonian of a physical system can be written as the sum of the kinetic and potential energy:

\begin{equation}
  \label{eq:hamiltonian}
  \mathcal{H}(\boldsymbol{q}, \boldsymbol{p}) = T(\boldsymbol{q}) + V(\boldsymbol{p})
\end{equation}
The HMC algorithm uses a combination of Gibbs sampling, Hamiltonian Dynamics and a Metropolis-Hastings step; hence, it is sometimes called Hybrid Monte Carlo. First the posterior distribution is augmented with an additional momentum parameter $\phi$. This is an auxiliary parameter which is not of direct interest when calculating the parameter posterior distribution. The parameters, $\psi$ correspond to the position in Hamilton's equations. We take the kinetic energy to be $T(\phi) = \frac{1}{2}\phi^T\phi$, assuming a unit particle mass. The joint density of the position and momentum can then be written as:
\begin{equation}
  \label{eq:3}
  p(\psi, \phi) \propto \exp \left\{ \log p(\psi|y) - \frac{1}{2}\phi^T\phi \right\},
\end{equation}
where $\log p(\psi|y)$ is the log of the target posterior distribution written up to a
normalising constant and represents the negative potential energy. The kinetic
energy is the kernel of a standard multivariate Normal distribution, and the
identity covariance matrix can replaced by a tuning parameter, $\Sigma$ termed
the mass matrix, from analytical mechanics~\citep{betancourt2017conceptual}. Hamilton's
equations are discretised in order to update the values of the static
parameters, $\psi$ (the position in Hamilton's equations) and the momentum
$\phi$. A special discretisation of Hamilton’s equations is used called a leapfrog step:

\begin{align*}
\phi_{t+\varepsilon/2} &= \phi_{t} + \frac{\varepsilon}{2} \nabla_\psi\log p(\psi_{t}|y), \\
\psi_{t+\varepsilon} &= \psi_{t} + \varepsilon \phi_{t+\varepsilon/2}, \\
  \phi_{t+\varepsilon} &= \phi_{t+\varepsilon/2} + \frac{\varepsilon}{2} \nabla_\psi\log p(\psi_{t+\varepsilon}|y). \numberthis \label{eqn:leapfrog}
\end{align*}
$\nabla_\psi\log p(\psi|y)$ is the gradient of the un-normalised
log-posterior distribution with respect to the parameters, $\psi$. $\varepsilon$ is a tuning parameter in the HMC
algorithm and represents the step-size of a leapfrog step. This leapfrog update is more accurate than a naive Euler discretisation, primarily due to the fact that it is volume-preserving (since each step is a shear). The volume-preservation is important for HMC, since it facilitates reversibility, and avoids the need to track Jacobians. The steps required to
perform HMC are summarised in Algorithm~\ref{alg:hmc}. The function
$\operatorname{leapfrogs}$ is a recursive function which applies
$\operatorname{leapfrog\_step}$ $L$ times without mutating state and returns the
updated position and momentum.

\begin{algorithm}
\SetAlgoLined
\KwResult{Return $\psi_i, i = 1,\dots,M$}
Given the prior distribution $p(\psi)$\;
The un-normalised the log-posterior distribution $\log p(\psi | y)$ and its
derivative $\nabla_\psi \log p(\psi|y)$\;
The tuning parameters $\varepsilon$, $L$\;
Initialise the parameters by sampling $\psi_0 \sim p(\psi)$\;
 \For{i in 1 to M}{
   Sample $\phi \sim \textrm{MVN}(0, \Sigma)$.\;
   $\psi^\star, \phi^\star = \operatorname{leapfrogs}(\psi^\star, \phi^\star, L, \varepsilon)$\;
   Calculate $\alpha = \log p(\psi^\star(L)|y) - \frac{1}{2}\phi^{\star T}(L)\Sigma^{-1}\phi^\star(L) - \log p(\psi_{i-1}|y) + \frac{1}{2}\phi^T\Sigma^{-1}\phi$\;
   Sample $u \sim U(0,1)$\;
   \uIf{$\log(u) < \alpha$}{
     Set $\psi_i := \psi^\star$
   }\uElse{
     Set $\psi_i := \psi_{i-1}$
   }
 }
 \SetKwFunction{FMain}{leapfrog\_step}
 \SetKwProg{Fn}{function}{:}{}
  \Fn{\FMain{$\psi$, $\phi$, $\varepsilon$}}{
    $\tilde{\phi} = \phi + \frac{\varepsilon}{2} \nabla_\psi \log p(\psi|y)$\;
    $\tilde{\psi} = \psi + \varepsilon \Sigma^{-1}\tilde{\phi}$\;
    \Return{$\tilde{\psi}, \tilde{\phi} + \frac{\varepsilon}{2} \nabla_\psi\log p(\tilde{\psi}|y)$\;}
  }
 \SetKwFunction{FMain}{leapfrogs}
 \SetKwProg{Fn}{function}{:}{}
  \Fn{\FMain{$\psi$, $\phi$, $\varepsilon$, $L$}}{
   \uIf{$L = 0$}{
     \Return{$\psi, \phi$}
   }\uElse{
     $\tilde{\psi}, \tilde{\phi} = \operatorname{leapfrog\_step}(\psi, \phi, \varepsilon)$\;
     $\operatorname{leapfrogs}(\tilde{\psi}, \tilde{\phi}, \varepsilon, L - 1)$
   }
 }
 \caption{Hamiltonian Monte Carlo \label{alg:hmc}}
\end{algorithm}

For differentiable targets, HMC can be shown to exactly preserve the required distribution, and is typically more efficient then random walk Metropolis-Hastings schemes.
The optimal acceptance rate for HMC is approximately 0.65~\citep{neal2011mcmc}.
Selecting the tuning parameters corresponding to the leapfrog step size
$\varepsilon$ and the number of leapfrog steps $L$ is typically done using short
pilot runs of the chain and targeting an acceptance rate of 0.65. Efforts have
been made to automate selection of tuning parameters in the HMC algorithm and
have led to the No-U-turn sampler which performs the optimum number of leapfrog
steps~\citep{hoffman2014} and empirical HMC~\citep{wu2018faster}. In both cases the optimum number
of leapfrog steps are chosen such that the discretised steps in the posterior
distribution does not make an U-turn by first heading away from the previously
accepted parameter value, then turning back in the direction of the previous
parameter when the value of the gradient changes.

\section{Automatic Differentiation}
\label{sec:forward-ad}

The need to calculate gradients in many inference and optimisation algorithms
such as Hamiltonian Monte Carlo (HMC) and variational
inference~\citep{kucukelbir2017automatic} has led to a renewed interest in automatic differentiation (AD). AD is a way
of calculating derivatives of functions whilst at the same time evaluating them.
This is not numerical differentiation or symbolic differentiation but rather exact
differentiation which returns the value of a derivative at a
point~\citep{wang2018demystifying}. It is straightforward to implement forward
mode automatic differentiation in an FP language. However the number of
computations performed using forward mode AD depends on the dimension of the
input space, and hence does not scale well to parameter inference in models
involving a large number of parameters. On the other hand the number of
computations required for reverse mode automatic differentiation scales with the
dimension of the output dimension. In Bayesian inference and HMC in particular
the gradient of the log-posterior with respect to the free-parameters is
required, this is a function from $\mathbb{R}^n \rightarrow \mathbb{R}$, where
$n$ is the dimension of the parameter space, hence
reverse mode AD is typically more efficient.

Hamiltonian Monte Carlo was introduced in section~\ref{sec:hmc}, along
with some implementation difficulties, such as choosing the leapfrog step size, number of leapfrog steps and deriving the partial derivatives required for the
proposal. Automatic differentiation can be used to calculate the
exact derivatives needed when performing the leapfrog step of HMC.

Dual numbers can be used in order to calculate derivatives (of univariate functions) automatically and exactly. Each real number has a corresponding dual number, which is the number, $y \in \mathbb{R}$ plus a small innovation, $\varepsilon$ such that $\varepsilon^2 = 0$.

Derivatives can be calculated by evaluating functions using the dual number, for
instance the function $f: \mathbb{R} \rightarrow \mathbb{R}$ defined by $f(x) =
x^2 + 2x + 5$, the derivative is $f^\prime(x) = 2x + 2$. In order to calculate
the derivative automatically using dual numbers, the function is evaluated using
the dual number equivalent to a chosen value of $x$ for instance $x = 5$, has
the dual number $5 + \varepsilon$ then

\begin{align*}
f(5 + \varepsilon) &= (5 + \varepsilon)^2 + 2(5 + \varepsilon) + 5, \\
                   &= 25 + 10\varepsilon + \varepsilon^2 + 10 + 2\varepsilon + 5, \\
                   &= 40 + 12\varepsilon.
\end{align*}
In this way, the evaluation of $f$ using the dual number has resulted in the
evaluation of $f(x)$, and $f^\prime(x)$ simultaneously with $f^\prime(x)$ being
given by the coefficient of $\varepsilon$. This is the essence of
forward mode automatic differentiation.

Automatic differentiation using dual
numbers is equivalent to applications of the chain rule:

\begin{equation}
  \label{eq:chain-rule}
  (f \circ g)^\prime = (f^\prime \circ g)g^\prime
\end{equation}
Combinations of primitive functions can be differentiated using repeated
applications of the chain rule. \lstinline{flatMap} can be defined for the
\lstinline{Dual} class which encapsulates the chain rule. This presentation of
forward mode automatic differentiation was first outlined by~\citep{welsh2018}.
\begin{lstlisting}
case class Dual[A: Numeric](real: A, eps: Double) {
  def flatMap[B: Numeric](f: A => Dual[B]): Dual[B] =
    val x = f(real)
    val nextEps = eps * x.eps
    Dual(x.real, nextEps)
}
\end{lstlisting}
The \lstinline{Numeric} context bound is required since the
primitive functions, such as multiplication, are defined explicitly in
terms of their real result and derivative; hence an
\lstinline{eps: Double} has to be multiplied by a \lstinline{real: A} when
applying the product rule for differentiation:

\begin{lstlisting}
def times(x: Dual[A], y: Dual[A])(implicit ev: ConvertableFrom[A]) =
  Dual[A](x.real * y.real, 
    x.eps * ev.toType[Double](y.real) + 
      y.eps * ev.toType[Double](x.real))
\end{lstlisting}

The derivatives of special functions can be written as:

\begin{lstlisting}
def sin[A: Numeric](a: Dual[A])(implicit ev: A =:= Double) =
  Dual(math.sin(real), math.cos(real) * eps)
def log[A: Numeric](a: Dual[A])(implicit ev: A =:= Double) =
  Dual(math.log(a.real), a.eps / a.real)
\end{lstlisting}
The derivatives of the special functions can also be written using \lstinline{flatMap}

\begin{lstlisting}
def sin(a: Dual[A])(implicit ev: A =:= Double) =
  a.flatMap(x => Dual(math.sin(x), math.cos(x)))
def log(a: Dual[A])(implicit ev: A =:= Double) =
  a.flatMap(x => Dual(math.log(real), 1 / real))
\end{lstlisting}

The implementation becomes more complex as functions of multiple arguments are
considered, since a separate $\varepsilon$ is required for the derivative with
respect to each argument~\citep{manzyuk2012simply}.

\begin{figure}
  \centering
  \includegraphics[width=0.5\textwidth]{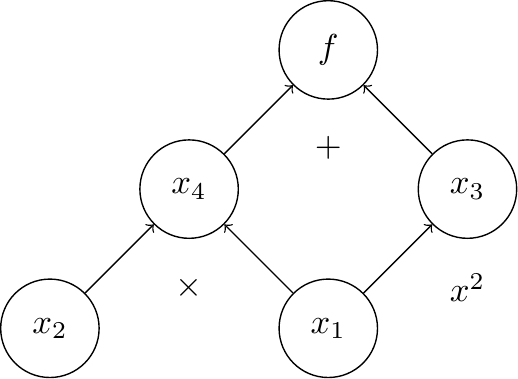}
  \caption[Decomposition of a function]{Computational graph showing the decomposition of $f(x_1, x_2) = x_1x_2 + x_1^2$ into primitive functions}
  \label{fig:ad-graph}
\end{figure}

Figure~\ref{fig:ad-graph} is a representation of a function with multiple arguments $f(x_1, x_2) = x_1x_2 + x_1^2$. The arrows represent function application, intermediate nodes represent the result of each primitive
function and are labelled $x_3$ and $x_4$. 

Differentiating the function with respect to all variables in the
graph using forward mode AD requires multiple traversals of
the graph. The derivative of a node, $x_k$ with parent nodes $Pa(x_k)$ with
respect to the input variable $x_1$ is derived from the chain rule

\begin{equation}
  \label{eq:chain-rule-graph}
  \frac{\partial x_k}{\partial x_1} = \sum_{j \in Pa(x_k)} \frac{\partial x_k}{\partial x_j} \frac{\partial x_j}{\partial x_1}.
\end{equation}

\begin{figure}
  \centering
  \includegraphics[width=0.5\textwidth]{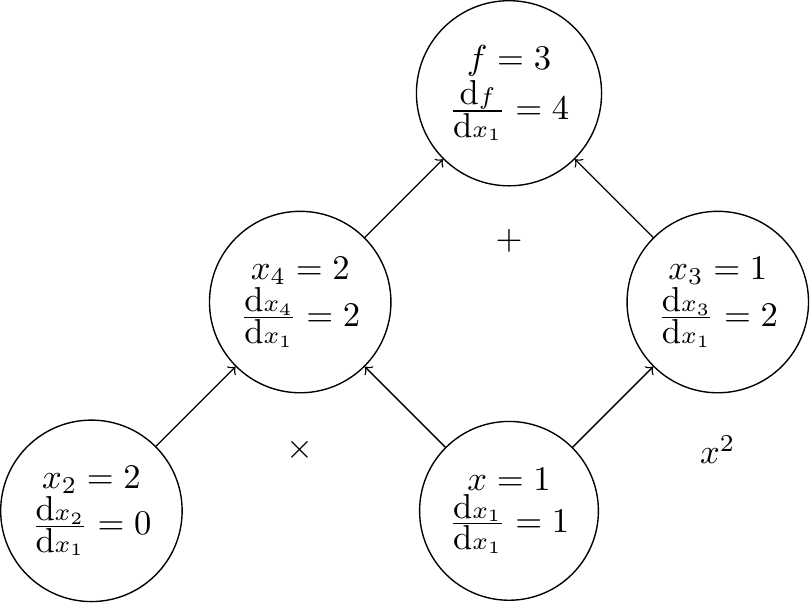}
  \caption[Graph of derivative]{Computational graph showing the applications of
    the chain rule required to calculate the partial derivative $\frac{\partial
      f}{\partial x_1}$.}
  \label{fig:forward-graph}
\end{figure}

Figure~\ref{fig:forward-graph} shows the forward pass required to
calculate the gradient $\frac{\partial f}{\partial x_1}\rvert_{x_1=1,x_2=2}$.
This single forward pass evaluates the function and the derivative with respect to $x_1$
given the known derivatives of primitive functions such as product and
power. A second pass has to be computed in order to determine the derivative
with respect to $x_2$. The gradient calculated by hand is, $\frac{\partial
  f}{\partial x_1} = x_2 + 2x_1$.
A method often used to implement AD for functions of multiple arguments $f: \mathbb{R}^n \rightarrow
\mathbb{R}$ is to track a vector of derivatives.

A minimal implementation of forward mode AD along with a test suite verifying
the algorithm using property based testing and a selection of different
functions is available at~\footnote{https://git.io/probprog}.

\subsection{Reverse Mode Automatic Differentiation}
\label{sec:reverse-ad}

Reverse mode automatic differentiation is typically faster than forward mode AD
when functions have a larger input space than output space, ie. $f: \mathbb{R}^n
\rightarrow \mathbb{R}^m$ where $n > m$. This is true for Bayesian inference
algorithms which require the gradient of the un-normalised log-posterior such as
Hamiltonian Monte Carlo or Metropolis-adjusted Langevin
algorithm~\citep{roberts1998optimal}. The un-normalised log-posterior is a
function from the parameters of dimension $n$ to a single real number. 

\begin{figure}
  \centering
  \includegraphics[width=0.5\textwidth]{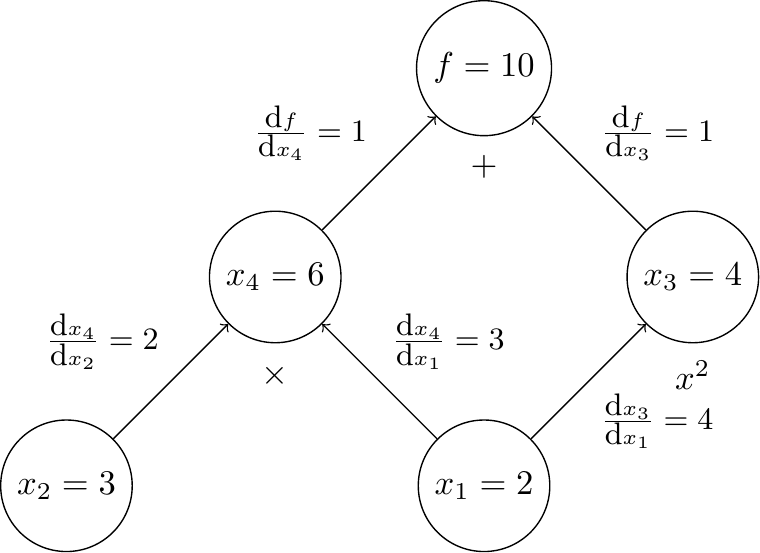}
  \caption[Adjoint graph for automatic differentiation]{Graph with derivatives of each input variable calculated on the nodes
    of the adjoint graph}
  \label{fig:auto-diff-adjoint}
\end{figure}

In order to differentiate a function using reverse mode automatic
differentiation the steps are similar to forward mode differentiation, the derivatives of simple functions are defined. The function is then decomposed into
its constituent primitive functions. The derivative of with respect to each node
of the adjoint graph is calculated at $x_1 = 2$, $x_2 = 3$ in figure~\ref{fig:auto-diff-adjoint}.

\begin{figure}
  \centering
  \includegraphics[width=0.5\textwidth]{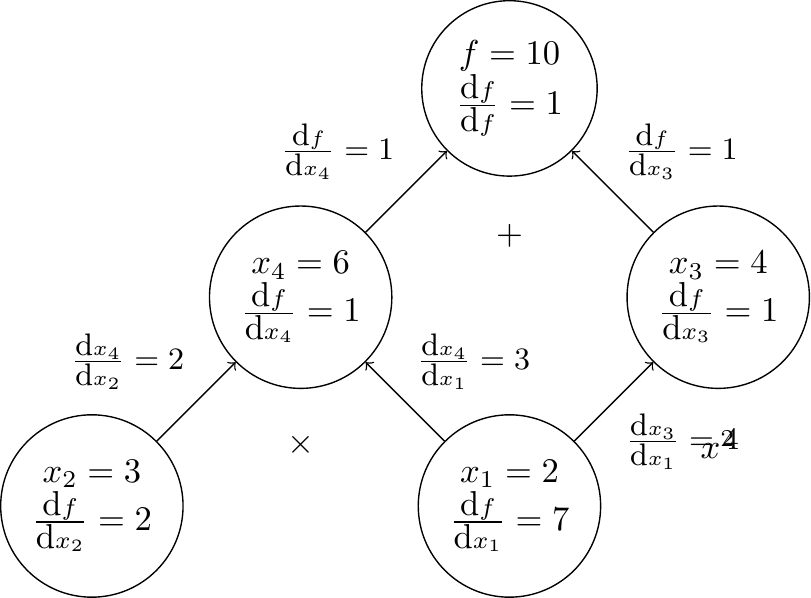}
  \caption{Reverse pass through the graph to calculate the derivative of $f$
    with respect to all the input variables}
  \label{fig:ad-reverse}
\end{figure}

Then proceeding in reverse through the graph to calculate the derivative of the
function with respect to the input variables:

\begin{equation*}
  \frac{\partial f}{\partial x_k} = \sum_{j \in Ch(x_k)} \frac{\partial x_j}{\partial x_k}\frac{\partial f}{\partial x_j}
\end{equation*}
Where $Ch(x_k)$ represent the set of children of node $x_k$.
Figure~\ref{fig:ad-reverse} shows the reverse sweep through the computation
graph required to calculate the derivative with respect to all of the arguments
of the function.

Automatic differentiation can be a part of a monadic
probabilistic program which allows distributions to be composed into
hierarchical models. Reverse mode automatic differentiation can be implemented
as a monad using continuations~\footnote{https://na.scaladays.org/schedule/differentiable-functional-programming}.

\section{Architecture of a Probabilistic Program}
\label{sec:architecture}

The Scala library Rainier~\citep{rainier2018} is an example of a monadic probabilistic programming language. It can be used to build a compute
graph over the parameters of a model. The compute graph is essentially a
variadic function from many input parameters to an output. This function is a representation of the un-normalised log-posterior function and
returns both the evaluation of the function and the gradient. The compute graph
can be compiled to a function with signature \lstinline{Array[Double] => Double}
to evaluate the un-normalised log-posterior and \lstinline{Array[Double] => Array[Double]} for the corresponding gradient.

The casual user is not aware of the compute graph and simply constructs a model
by combining primitive probabilistic functions. A distribution is defined as a
\lstinline{trait}:

\begin{lstlisting}[caption={Abstract interfaces for a continuous distribution},label={lst:distributions}]
trait Continuous[A] {
  def logPdf(a: A): Double
  def gen: Rand[A]
  def param: RandomVariable[Real]
  def fit(ys: A*): RandomVariable[A]
}
\end{lstlisting}
A continuous distribution can
evaluate the log probability density function, \lstinline{logPdf} and be
transformed into a \lstinline{RandomVariable} monad for use as a parameter in a
composed model using \lstinline{param}. Discrete distributions can be defined similarly, except the log
probability mass function is defined: \lstinline{logPmf} and discrete
distributions can not be used as a parameter in HMC since it can not be
differentiated and hence the \lstinline{param} method is not defined for
discrete distributions. The \lstinline{fit} method accepts as an argument a
single observation, or multiple and allows the distribution to be used as the
likelihood for the observation. Each interface has a method, \lstinline{gen} which can
transform the distribution to a sampling based distribution, called a
\lstinline{Generator} in Rainier and sometimes referred to as \lstinline{Rand}. \lstinline{Rand} is also a monad:

\begin{lstlisting}
trait Rand[A] { self =>
  def draw(): A
  def map[B](f: A => B): Rand[B] = new Rand[B] {
    override def draw = f(self.draw)
  }
  def flatMap[B](f: A => Rand[B]): Rand[B] = new Rand[B] {
    override def draw = f(self.draw).draw
  }
}
\end{lstlisting}
and has similar semantics to the \lstinline{Rand} monad discussed earlier in the context of pure functional random number generation.
Then the only function which remains abstract and must be defined for a new
\lstinline{Rand} is \lstinline{draw}, hence this monad has no information about
the \lstinline{logPmf} or \lstinline{logPdf} functions. The functions
\lstinline{flatMap} and \lstinline{pure} are the two functions required to
define a monad, if the natural transformations defined on \lstinline{Rand}
satisfy the monad laws then \lstinline{Rand} is a monad. 

A model to infer the posterior distribution for the probability of
heads on a coin from ten coin flips can be written as

\begin{lstlisting}[caption={Rainier probabilistic program expressing a coin flip example with a beta prior distribution for the probability of heads},label={lst:prob-coin-rainier}]
val model = for {
  p <- Beta(3, 3).param
  _ <- Binomial(p, 10).fit(6)
} yield p
\end{lstlisting}
Using \lstinline{.param} to indicate the bounded parameter $p$ has a beta prior
distribution with bounded support on $[0, 1]$, then the discrete binomial
distribution is used as the likelihood by calling the method \lstinline{fit} with
the total trials $n = 10$ and observed number of heads $y = 6$. Each model
defined using this monadic syntax can be sampled from using a
\lstinline{Sampler}. The \lstinline{Sampler} depends on the compute graph which can derive the un-normalised log-posterior and its gradient.

\section{Example Probabilistic Programs}
\label{sec:example-pp}

This section contains some example programs written using the Rainier
DSL. The models are intentionally simple, as they are intended to present just the most basic concepts. Some areas in which assumptions of these
simple models can be relaxed are highlighted --- in order to improve model fit.
This emphasises the flexibility of probabilistic programming.

For each of the examples the HMC algorithm with dual averaging and $L = 5$
leapfrog steps was used with 10,000
iterations for the warm-up run during which the initial leapfrog step size is
determined followed by 50,000 sampling iterations with a thinning factor of 5.

\subsection{Linear Model}
\label{sec:lm-pp}

\begin{figure}
  \centering
    \subfloat[]{
    \includegraphics[width=0.4\textwidth]{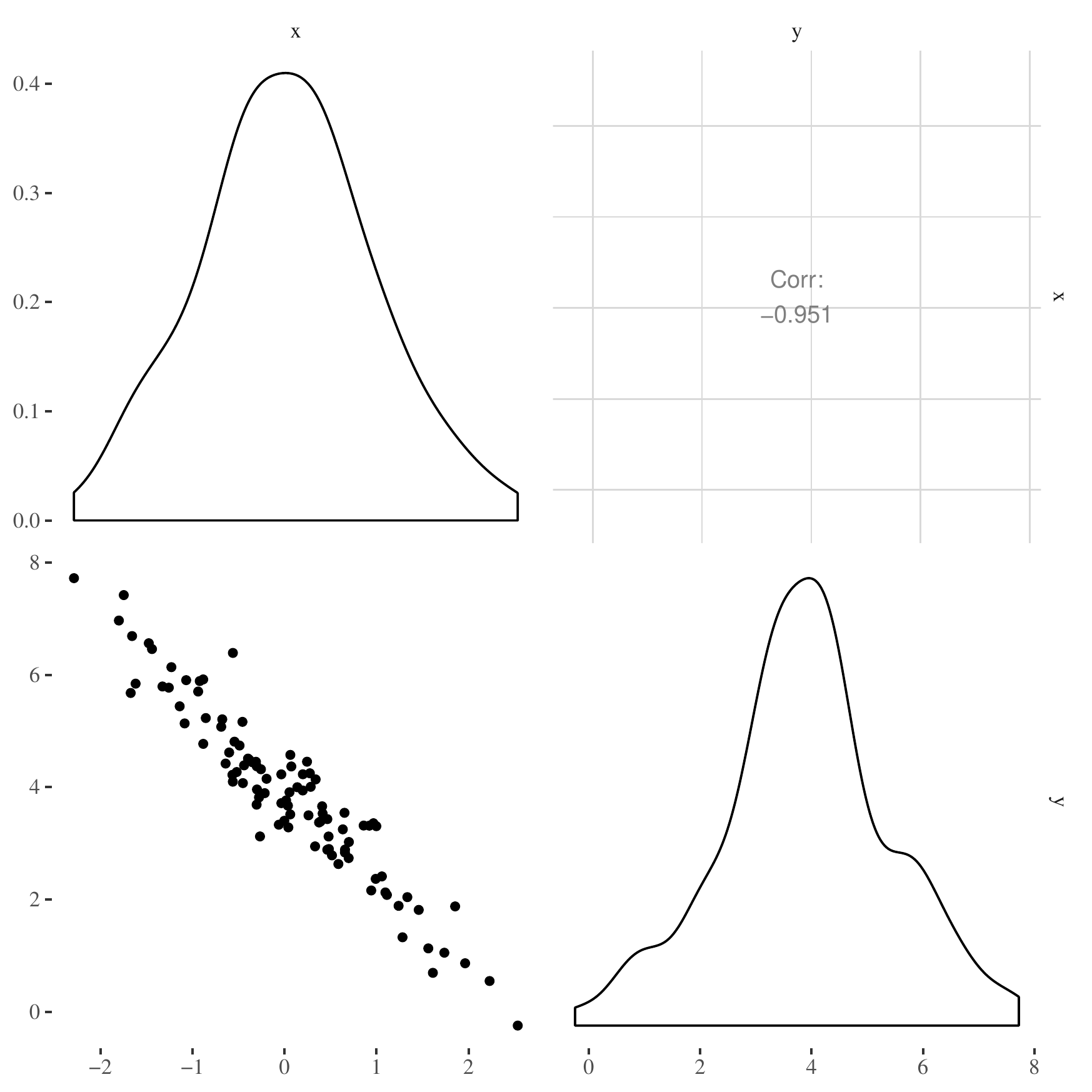}}\qquad
  \subfloat[]{\includegraphics[width=0.4\textwidth]{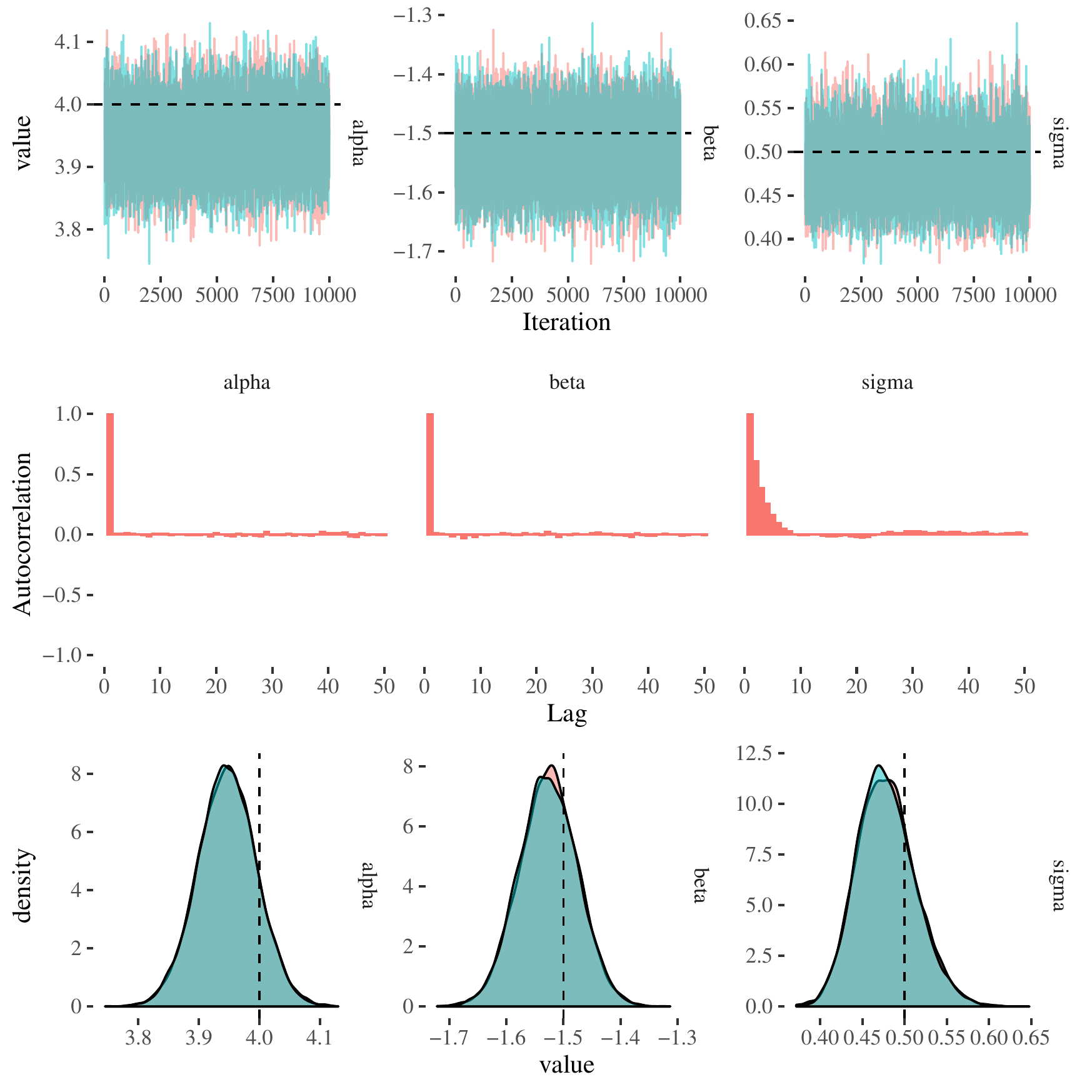}}
  \caption{(a) Simulated data from a linear model with
    $\beta_0 = 4.0, \beta_1 = -1.5, \sigma = 0.5$ and covariates
    simulated independently from a standard Normal distribution. (b) Parameter posterior diagnostic plots for
    the linear model using the HMC kernel and simulated data, the actual
    values of the parameters are plotted using the dashed lines (Top) Traceplots
  (Middle) Autocorrelation plots (Bottom) Marginal Densities}
  \label{fig:lm}
\end{figure}

Linear models are statistical models specifying a relationship between
covariates, $X \in \mathbb{R}^{n \times p}$ and a univariate outcome $Y \in
\mathbb{R}^n$ for each example
via the coefficient $\beta$, a $(p + 1)$-vector. Each row of the outcome
matrix $Y$ is denoted as $y_i$ and is related to each row of the covariate
matrix, $x_i$ by the coefficient matrix $\beta$. The covariate matrix has a
column vector of ones prepended as the first column which represents the intercept. The observations have
independent normally distributed noise with equal variance:

\begin{align*}
  y_i &= \beta^T x_i + \varepsilon_i, \qquad \varepsilon_i \sim \mathcal{N}(0,
  \sigma), \numberthis  \label{eq:lm}\\
  \beta &\sim \textrm{MVN}(\mu_\beta, \Sigma_\beta^2), \\
  \sigma &\sim \textrm{Exponential}(\lambda_\sigma).
\end{align*}
The model assumptions can be relaxed, but then model interpretation can become more
challenging. 1,000 observations from this model are simulated with $\beta_0 =
4.0, \beta_1 = -1.5, \sigma = 0.5$ and the covariates simulated
from the standard Normal distribution. The bivariate relationships of the
simulated data is plotted in Figure~\ref{fig:lm} (a). The figures were produced
using R~\citep{r2019} and ggplot2~\citep{hadley2016}.

Listing~\ref{lst:lm} shows a probabilistic program for a linear
model with covariates \lstinline{x} and outcome \lstinline{y}. Each parameter is
added to the model using the \lstinline{.param} notation and the $p = 1$
covariates are related to each scalar outcome using
\lstinline{Predictor.from}. The linear model is expressed as a function which
can be re-used in more complex hierarchical models, such as the random effects model
presented in section~\ref{sec:random-effects}.

\begin{lstlisting}[caption={Simple linear model with one covariate and an intercept with a Normal observation model and fully specified prior distributions\label{lst:lm}},numbers=left,stepnumber=1]
def linearModel(pa: Real, pas: Real, pb: Real, pbs: Real,
                sigma: Real, data: Vector[(Double, Double)]) = for {
  alpha <- Normal(pa, pas).param
  beta <- Normal(pb, pbs).param
  _ <- Predictor[Double].from { x =>
      Normal(alpha + beta * x, sigma)
    }
    .fit(data)
} yield Map("alpha" -> alpha, "beta" -> beta, "sigma" -> sigma)

val model = for {
  sigma <- Exponential(3.0).param
  params <- linearModel(0.0, 0.01, 0.0, 0.01, sigma, data)
} yield params
\end{lstlisting}
Figure~\ref{fig:lm} (b) shows posterior
inferences using simulated data with $\beta_0 = 4.0, \beta_1 = -1.5, \sigma = 0.5$ and the covariates simulated from a standard Normal
distribution. 

It is straightforward to change the observation distribution of this model, for
instance to a generalised linear model with a Poisson observation distribution
by changing line 7 to:

\begin{lstlisting}
Poisson((alpha + beta * x).exp)
\end{lstlisting}
In addition the standard deviation parameter \lstinline{sigma} is no longer
required when using a Poisson observation distribution. Any suitable prior
distributions can be specified for the parameters; they do not have to be conditionally conjugate as in Gibbs sampling.

\subsection{Mixture Model}
\label{sec:mixture-model}

In a mixture model the observed data is assumed to arise from a finite
mixture of independent distributions. The mixture model considered here is a
mixture of $m = 3$ Normal distributions with different mean values,  $\mu_k, k =
1,\dots,m$ and a common variance.

\begin{align*}
  y_i &\sim \mathcal{N}(\mu_k, \sigma), \\
  k &\sim \mathcal{F}(\theta), \\
  \theta &\sim \textrm{Dir}(\pmb\alpha_\theta), \\
  \mu_k &\sim \mathcal{N}(\mu_{\mu_k}, \sigma^2_{\mu_k}), \\
  \sigma &\sim \textrm{Exponential}(\alpha_\sigma, \beta_\sigma).
\end{align*}
where each index $k$ is drawn from a discrete distribution $\mathcal{F}$ with
probabilities $\theta$. The likelihood of the mixture model can be written as:

\begin{equation}
  \label{eq:19}
  p(y|\theta, \mu, \sigma) = \sum_{k=1}^m \theta_k \mathcal{N}(y; \mu_k, \sigma).
\end{equation}
This mixture distribution has a smooth log-density and hence can be
differentiated and inference can be performed using HMC. In addition the mixture
distribution is already implemented in Rainier. The mixture model can be defined as:

\begin{lstlisting}[caption={Rainier model for a Gaussian mixture model with $m = 3$ components.}]
def normalise(alphas: Seq[Real]) = {
  val total = alphas.reduce(_ + _)
  alphas.map(a => a / total)
}

val model = for {
  unnormThetas <- RandomVariable.traverse(Vector.fill(3)(Gamma(3.0, 1.0).param))
  thetas = normalise(unnormThetas)
  mus <- RandomVariable.traverse(Vector.fill(3)(Normal(0, 1).param))
  sigma <- Exponential(3.0).param
  dists = mus.
    map(mu => Normal(mu, sigma)
  components: Map[Continuous, Real] = dists.zip(thetas).toMap
  _ <- Mixture(components).fit(ys)
} yield thetas ++ mus ++ sigma
\end{lstlisting}

\begin{figure}
  \centering
  \subfloat[]{
    \includegraphics[width=0.4\textwidth]{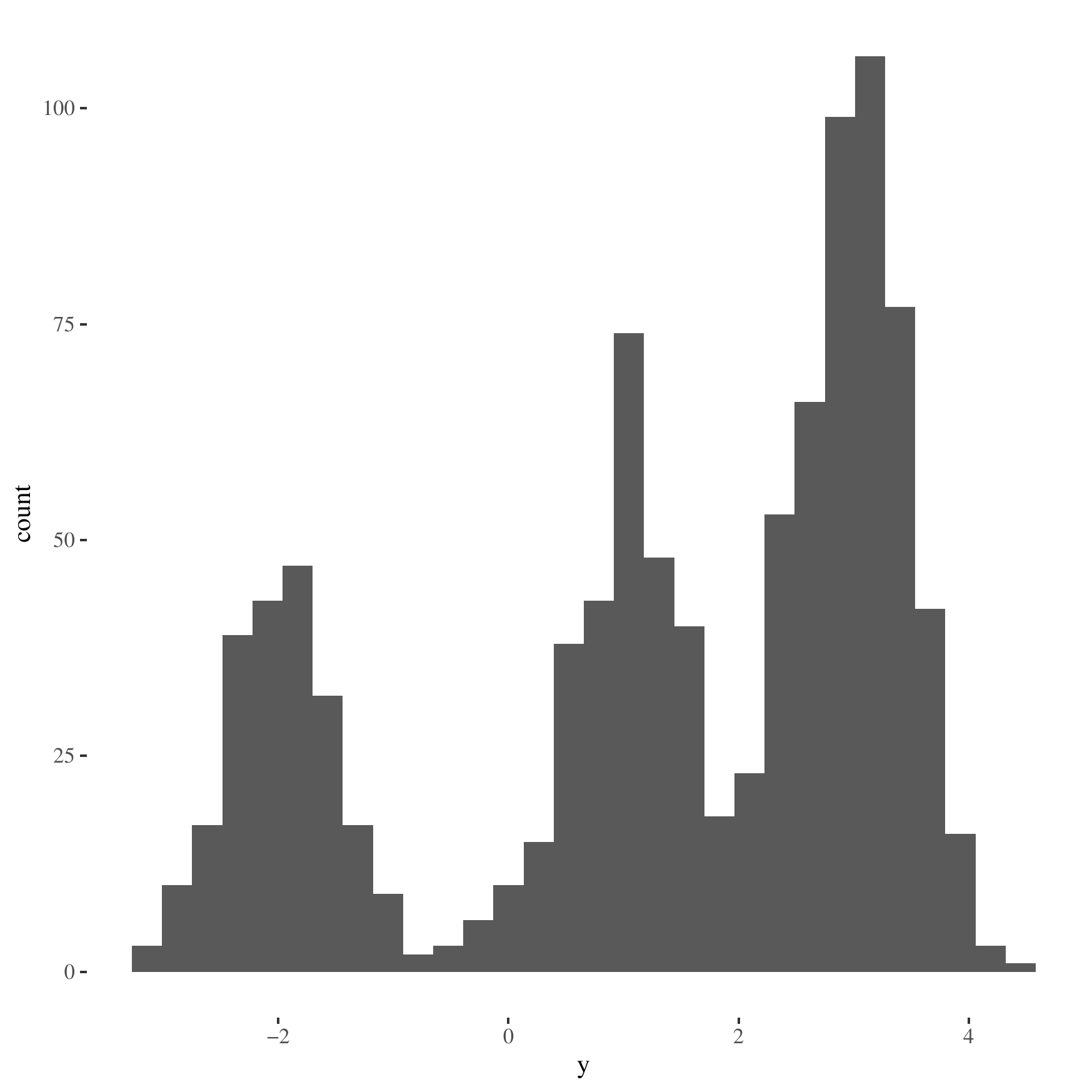}}\qquad
  \subfloat[]{\includegraphics[width=0.4\textwidth]{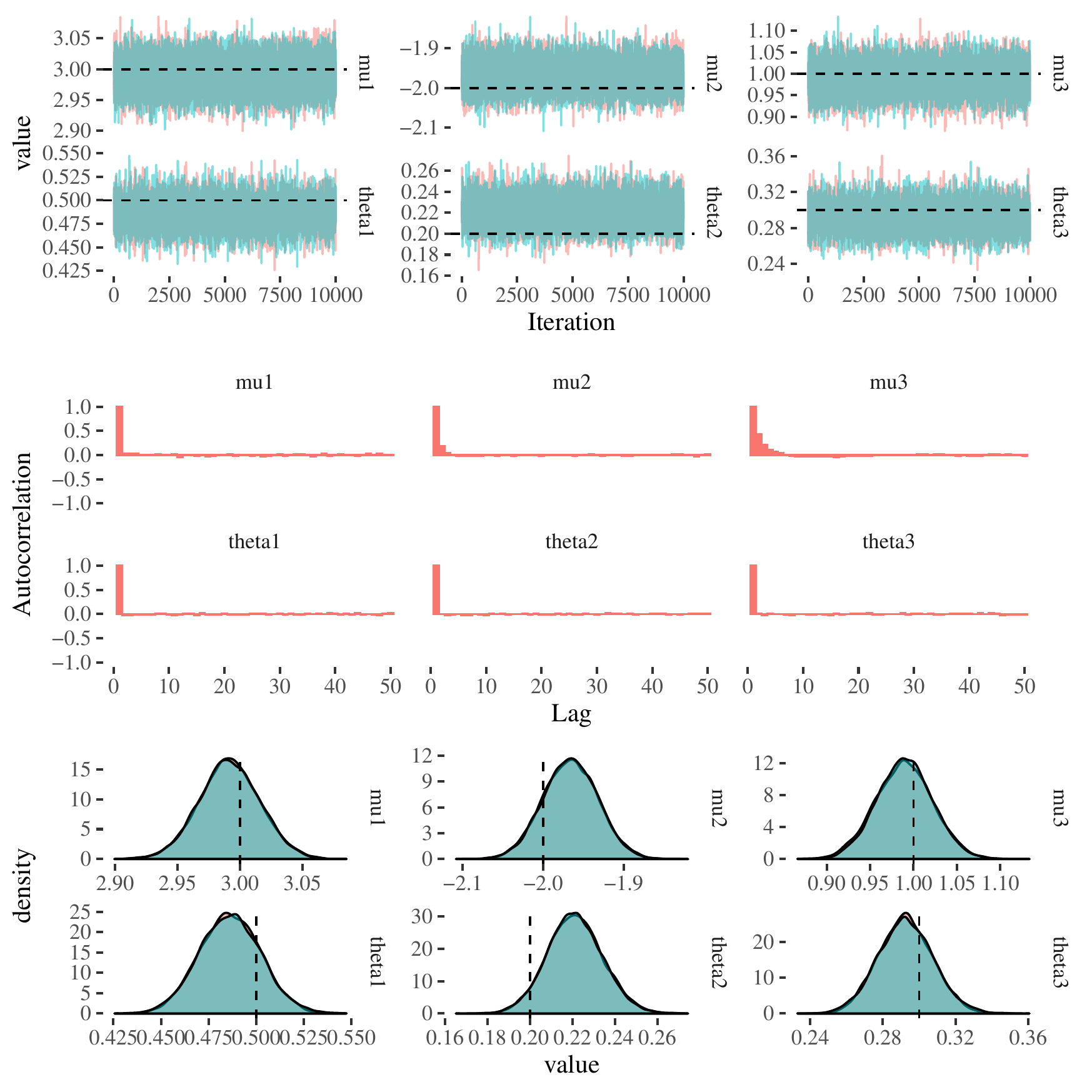}}
  \caption[Mixture Model simulation]{(a) 10,000 simulations from a three component
    Normal mixture model with mean values $\mu = \{-2.0, 1.0, 3.0\}$, common
    variance $\sigma = 0.5$ and proportions $\theta = \{0.3, 0.2, 0.5\}$ (b)
    Parameter posterior marginal density plots for the mixture
    model parameters, the parameter values used to simulate the data are vertical dashed lines}
  \label{fig:mixture-model}
\end{figure}

The probability of each mixing component is $\theta_i$, $i = 1,2,3$. The mixing
components must be greater than zero and sum-to-one, they are drawn from a
Dirichlet distribution by drawing each component from a Gamma
distribution with scale, $\theta = 1$ then normalising the values. Figure~\ref{fig:mixture-model} (a) shows a simulation from the mixture model and
(b) shows the posterior diagnostics for the means and mixing components. 

\subsection{Random Effects Model}
\label{sec:random-effects}

Hierarchical models give an opportunity to emphasise the manipulation of probabilistic programs as
values in the embedded DSL. This model is a random effects model with $k$
classes each with $n$ observations. Each class is modelled using an independent
linear regression.
 
\begin{align*}
  Y_{ij} &= \alpha_i + \beta_i x_j + \varepsilon_{ij}, &\varepsilon_{ij} \sim \mathcal{N}(0, \sigma^2), \\
  \alpha_i &\sim \mathcal{N}(\alpha_c, \sigma_a^{2}), \\
  \beta_i &\sim \mathcal{N}(\beta_c, \sigma_b^{2}).
\end{align*}
Each of the $i = 1,\dots,k$ classes have the same covariates $x_j, j = 1,\dots,n$ and have a linear relationship with
independent coefficients, $\alpha_i$ and $\beta_i$. The standard deviation of
the observation, $\sigma$ is assumed to be the same for each class and
observation. The coefficients of the covariates for all models are assumed to be Normally
distributed with the same mean and variance, this induces correlation between
the $k$ linear models. $\alpha_c, \sigma_a, \beta_c, \sigma_b$ and $\sigma$ are given the following prior distributions:

\begin{align*}
  \alpha_c &\sim \mathcal{N}(0, 10^6), \\
  \sigma_a &\sim \textrm{Gamma}(10^{-3}, 10^{3}), \\
  \beta_c &\sim \mathcal{N}(0, 10^6), \\
  \sigma_b &\sim \textrm{Gamma}(10^{-3}, 10^{3}), \\
  \sigma &\sim \textrm{Exponential}(3).
\end{align*}
The Gamma distribution is parameterised here using shape, $k$ and scale,
$\theta$ such that the mean is $k \theta$.
Each observation is represented by a class containing the index of the class, the
covariate, $x$, and the observation, $y$:

\begin{lstlisting}
case class Observation(
  id: Int,
  x: Double,
  y: Double
)
\end{lstlisting}
Then the prior distributions are translated to Rainier, note that the Gamma
distribution is parameterised in terms of shape, $k$ and scale, $\theta$ and the
Normal distribution by its mean, $\mu$ and standard deviation $\sigma$.

\begin{lstlisting}
val prior = for {
  alphaC <- Normal(0.0, 1000).param
  alphaSigma <- Gamma(0.001, 1000).param
  betaC <- Normal(0.0, 1000).param
  betaSigma <- Gamma(0.001, 1000).param
  sigma <- Exponential(3).param
} yield (alphaC, alphaSigma, betaC, betaSigma, sigma)
\end{lstlisting}
A function to fit a single linear regression for the
$i^{\textrm{th}}$ class can be re-used from the function \lstinline{linearModel} presented
in section~\ref{sec:lm-pp}.

\lstinline{linearModel} returns a \lstinline{RandomVariable} monad which is its own
probabilistic program, but this program only specifies the model for a
single class. The collection of observations, \lstinline{x} and \lstinline{y}
for a single class can be grouped by the \lstinline{id} of the class using \lstinline{groupBy} and a linear regression can be fit for each class using \lstinline{map}. When
the function \lstinline{groupBy} is called on the collection of observations, it returns
a \lstinline{Map[Int, Iterable[Observation]]}, the key of the map is an integer
representing the \lstinline{id} of each class and the  \lstinline{Iterable} in the
value of the \lstinline{Map} corresponds to the observations for each class. The
application of \lstinline{linearModel} to each element of the \lstinline{Map} returns an
\newline\lstinline{Iterable[RandomVariable[(Double, Double)]} which is converted to a
\lstinline{Vector} before applying the function
\lstinline{RandomVariable.traverse}. \lstinline{traverse} ``reverses'' the order of the effects
to return a \lstinline{RandomVariable[Vector[(Double, Double)]]} which is a model
and can be sampled from.

\begin{lstlisting}
val model: RandomVariable[Map[String, Real]] = for {
  (ac, sa, bc, sb, sigma) <- prior
  _ <- RandomVariable.traverse(
    observations.
      groupBy(_.id).
      map { case (id, obs) =>
        linearModel(ac, sa, bc, sb, sigma, obs.map(ys => (ys.x, ys.y)))
      }.toVector)
} yield (ac, sa, bc, sb, sigma)
\end{lstlisting}

This implementation emphasises the compositionality of probabilistic programs
embedded in a host language. The higher order functions, \lstinline{groupBy} and
\lstinline{map} are familiar to any functional programmer and can be used to
combine simple models into complex hierarchical models.

\begin{figure}
  \centering
  \includegraphics[width=0.75\textwidth]{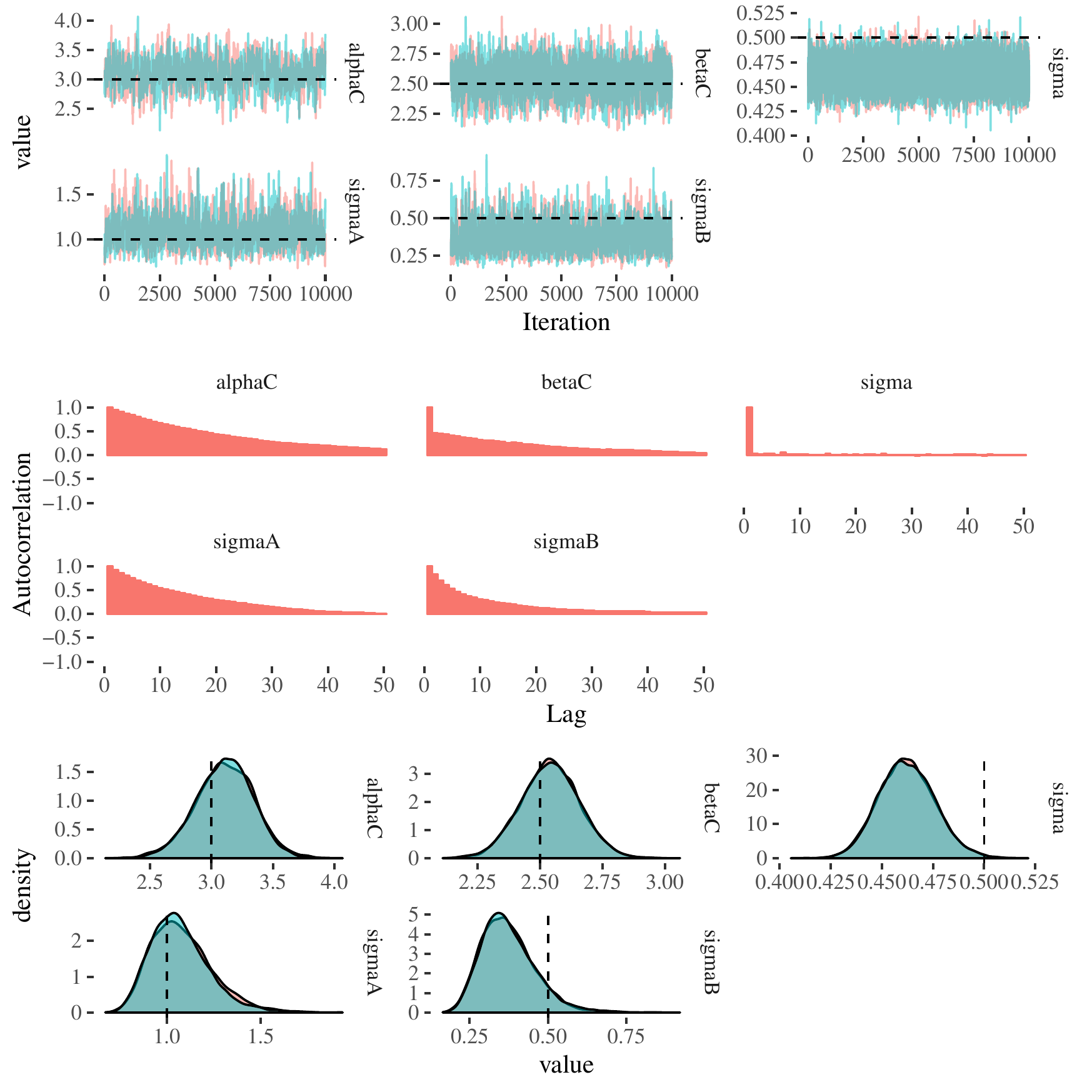}
  \caption[Posterior random effects]{Diagnostic plots for the hierarchical random effects
    model with the values used to simulate the overlaid using dashed lines (Top)
    Traceplots (Middle) Autocorrelation (Bottom) Empirical marginal densities}
  \label{fig:random-effects}
\end{figure}

The diagnostics of the draws from the posterior distribution obtained using HMC
on the random effects hierarchical example are presented in
figure~\ref{fig:random-effects}. The actual values used to simulate the data are
plotted using dashed lines.

\section{Conclusion}
\label{sec:conclusion}

Probabilistic programming aims to unify Bayesian inference with general purpose
programming languages in order to simplify model exploration and inference. This
allows practitioners to develop novel statistical models with minimal
consideration for the underlying inference algorithms. In addition, multiple
inference algorithms can be compared without re-writing the model code.

It has been shown that functional programming in Scala is a suitable, powerful language
for developing a probabilistic programming language as an embedded DSL. Static typing reduces
difficult to find runtime errors and ensures that refactoring code is
straightforward with the help of the compiler. This means code is not
over-engineered from the start of project as changes can be made to any size
code base as and when the specification changes. Static typing does not increase
the size of the code base since well defined abstractions from category theory
allow for polymorphic code with type safety. Additionally, Scala has local type
inference, allowing for return types which are obvious to be omitted from the
source code and inferred by the compiler, which is especially helpful when defining anonymous functions.

In order to maximise the utility of the statistical models, the DSL is embedded
in an existing general purpose programming language. Functional programming
languages have powerful, theoretically sound abstractions such as the free monad
and tagless final encodings which can be used to develop embedded DSLs for probabilistic programming. This means the same model code is used in production as
in development, reducing the possibility of bugs in production code leading to
incorrect decisions being made. The Rainier DSL provides a powerful modelling
language embedded in the Scala language. The `for'-comprehension used to build
models from probability distributions is familiar to
Scala programmers since many common data constructors form a monad. The monadic
syntax provided by `for' is reminiscent of the syntax in BUGS, JAGS and Stan
allowing experts in these languages to easily transition to using the embedded
DSL in Scala. Embedding the probabilistic programming DSL in a host functional
language not only renders trivial the problem of “compiling” the language to an
executable sampling algorithm, but also allows probabilistic programs to be
manipulated as regular values in the host language. This maximises the
opportunities for model composition, modular model development and code reuse.
This is exemplified in the random effects
example in section~\ref{sec:random-effects} .

The Rainier DSL
provides a powerful modelling language embedded in the Scala language. The for comprehension used to build
models from probability distributions is familiar to
Scala programmers since many common data constructors form a monad. The monadic
syntax provided by for is reminiscent of the syntax in BUGS, Jags and Stan
allowing experts in these languages to easily transition to using the embedded
DSL in Scala. This syntax is available with minimal effort when a monad is
defined by specifying the \lstinline{flatMap} and \lstinline{pure} functions. 

\section*{Acknowledgements}

JL is supported by the Engineering and Physical Sciences Research
Council, Centre for Doctoral Training in Cloud Computing for Big Data
(grant number EP/L015358/1) and Digital Catapult Teaching Grant Award
(KH153326).  DJW would like to thank the Isaac Newton Institute for
Mathematical Sciences for support and hospitality during the programme
Statistical Scalability when some work on this paper was
undertaken; EPSRC grant EP/R014604/1.
This work was also supported by The Alan Turing Institute under the EPSRC
grant EP/N510129/1.

\bibliography{bibliography}

\end{document}